\documentclass[11pt]{article}
\usepackage[utf8]{inputenc}
\usepackage{graphicx}
\usepackage{subfigure}

\usepackage{float}
\usepackage{wrapfig}
\usepackage{amsmath}
\usepackage{amssymb}
\usepackage{amsfonts}
\usepackage{hyperref}
\usepackage{bbm}
\usepackage[top=1in, bottom=1.25in, left=1.25in, right=1.25in]{geometry}

\newcommand{\RR}{\mathbb{R}}

\newcommand{\bin}{\beta_{\rm in}}	
\newcommand{\bout}{\beta_{\rm out}}

\newcommand{\binn}[1]{\beta_{{\rm in},{#1}}}	

\newcommand{\bincohh}[1]{\bar{\beta}_{{\rm in},#1}}
\newcommand{\z}{\alpha}			
\newcommand{\zz}[1]{\alpha_{#1}}

\providecommand{\keywords}[1]{\textbf{{Keywords:}} #1}

\begin{document}



\title{A Coherent Perceptron for All-Optical Learning}


\author{Nikolas Tezak\footnote{ntezak@stanford.edu}, Hideo Mabuchi \\
Edward L. Ginzton Laboratory, Stanford University\\
Stanford, CA 94305, USA}

\maketitle

\begin{abstract} 
We present nonlinear photonic circuit models for constructing programmable linear transformations and use these to realize a coherent Perceptron, i.e., an all-optical linear classifier capable of learning the classification boundary iteratively from training data through a coherent feedback rule. Through extensive semi-classical stochastic simulations we demonstrate that the device nearly attains the theoretical error bound for a model classification problem.
\end{abstract}








\section{Introduction} 
\label{sec:introduction}

Recent progress in integrated nanophotonic engineering \cite{Kippenberg2004KerrNonlinearity,Haye2007Optical,Levy2005Nanomagnetic,Razzari2009CmosCompatible,Englund2007Controlling,Fushman2008Controlled,Nozaki2010SubFemtojoule,Cohen2014Phonon,Vandoorne2014Experimental,Santori2014Quantum} has motivated follow-up proposals \cite{Mabuchi2011Nonlinear,Pavlichin2013Photonic} of nanophotonic circuits for all-optical information processing. While most of these focus on implementations of digital logic, we present here an approach to all-optical analog, \emph{neuromorphic} computation and propose design schemes for a set of devices to be used as building blocks for large scale circuits.

Optical computation has been a long-time goal \cite{Abraham1982Optical,Smith1984Optical}, with research interest surging regularly after new engineering capabilities are attained \cite{Miller1997Physical,Miller2010Are}, but so far the parallel progress and momentum of CMOS based integrated electronics has outperformed all-optical devices.

In recent years we have seen rapid progress in the domain of machine learning, and artificial intelligence in general. Although most current `big data'-applications are realized on digital computing architectures, there is now an increasing amount of computation done in specialized hardware such as GPUs. Specialized analog computational devices for solving specific subproblems more efficiently than possible with either GPUs or general purpose computers are being considered or already implemented by companies such as IBM, Google and HP and in academia, as well. \cite{Ananthanarayanan2009Cat,Neven2014Hardware,Strukov2008Missing,Wang2013Coherent}
Specifically in the field of neuromorphic computation, there has been impressive progress on CMOS based analog computation platforms \cite{Choudhary2012Silicon,Cassidy2014RealTime}. 

Several neuromorphic approaches to use complex nonlinear optical systems for machine learning applications have recently been proposed \cite{Duport2012AllOptical,Vandoorne2011Parallel,Vaerenbergh2012Cascadable,Dejonckheere2014AllOptical} and some initial schemes have been implemented \cite{Larger2012Photonic,Vandoorne2014Experimental}. So far, however, all of these `optical reservoir computers' have still required digital computers to prepare the inputs and process the output of these devices with the optical systems only being employed as static nonlinear mappings for dimensional lifting to a high dimensional feature space \cite{Cover1965Geometrical}, in which one then applies straightforward linear regression or classification for learning an input-output map. \cite{Verstraeten2010Reservoir}

In this work, we address how the final stage of such a system, i.e., the linear classifier could be realized all-optically. We provide a universal scheme, i.e., independent of which particular kind of optical nonlinearity is employed, for constructing \emph{tunable} all-optical, phase-sensitive amplifiers and then outline how these can be combined with self-oscillating systems to realize an optical amplifier with \emph{programmable} gain, i.e., where the gain can be set once and is then fixed subsequently.

Using these as building blocks we construct an all-optical \emph{perceptron} \cite{Rosenblatt1957PerceptronA,Rosenblatt1958Perceptron}, a system that can classify multi-dimensional input data and, using pre-classified training data learn the correct classification boundary `on-line', i.e., incrementally. The perceptron can be seen as a highly simplified model of a neuron. 
While the idea of all-optical neural networks has been proposed before \cite{Miller1993Novel} and an impressive scheme using electronic, measurement-based feedback for spiking optical signals has been realized \cite{Fok2013Pulse}, to our knowledge, we offer the first complete description for how the synaptic weights can be stored in an optical memory and programmed via feedback.

The physical models underlying the employed circuit components are high intrinsic-$Q$ optical resonators with strong optical nonlinearities. For theoretical simplicity we assume resonators with either a $\chi_2$ or a $\chi_3$ nonlinearity, but the design can be adapted to depend on only one of these two or alternative nonlinearities such as those based on free carrier effects or optomechanical interactions.

The strength of the optical nonlinearity and the achievable $Q$-factors of the optical resonators determine the overall power scale and rate at which a real physical device could operate. Both a stronger nonlinearity and higher $Q$ allow operating at lower overall power.

We present numerical simulations of the system dynamics based on the semi-classical Wigner-approximation to the full coherent quantum dynamics presented in \cite{Santori2014Quantum}. For photon numbers as low as ($\sim 10-20$) this approximation allows us to accurately model the effect of optical quantum shot noise even in large-scale circuits.

In the limit of both very high $Q$ and very strong nonlinearity, we expect quantum effects to become significant as entanglement can arise between the field modes of physically separated resonators. In the appendix, we provide full quantum models for all basic components of our circuit. The possibility of a quantum speedup is being addressed in ongoing work.
Recently, D-Wave Systems has generated a lot of interest in their own superconducting qubit based quantum annealer. Although the exact benefits of quantum dynamics in their machines has not been conclusively established \cite{Boixo2014Evidence}, recent results analyzing the role of tunneling in a quantum annealer \cite{Boixo2014Computational} are intriguing and suggest that quantum effects can be harnessed in computational devices that are not unitary quantum computers.

\subsection{The Perceptron algorithm} 
\label{sub:review_of_the_perceptron_algorithm}

The perceptron is a machine learning algorithm that maps an input $x\in\RR^n$ to a single binary class label $\hat{y}_w[x]\in\{0, 1\}$. Binary classifiers generally operate by dividing the input space into two disjoint sets and identifying these with the class labels.
The perceptron is a linear classifier, meaning that the surface separating the two class label sets is a linear space, a hyperplane, and its output is computed simply by applying a step function $\theta(u):=\mathbbm{1}_{u \ge 0}$ to the inner product of a single data point $x$ with a fixed \emph{weight vector} $w$:
\begin{align}
    \hat{y}_w[x] := \theta(w^Tx) = \begin{cases} 1 \text{ for } w^Tx \ge 0, \\ 0 \text{ otherwise.}\end{cases}
\end{align}
Geometrically, the weight vector $w$ parametrizes the hyperplane $\{z\in\RR^n:\; w^T z=0\}$ that forms the decision boundary.

In the above parametrization the decision boundary always contains the origin $z=0$, but the more general case of an affine decision boundary $\{\tilde{z}\in\RR^n:\; \tilde{w}^T \tilde{z} = b\}$ can be obtained by extending the input vector by a constant $z = (\tilde{z}^T, 1)^T\in\RR^{n+1}$ and similarly defining an extended weight vector $w=(\tilde{w}^T, -b)^T$.

The perceptron converges in a finite number of steps for all linearly separable problems \cite{Rosenblatt1957PerceptronA} by randomly iterating over a set of pre-classified training data $\{(y^{(j)},x^{(j)}) \in \{0, 1\} \otimes \RR^n,\; j=1, 2,\dots, M\}$
and imparting a small weight correction $w\to w + \Delta w$ for each falsely classified training example $x^{(j)}$
\begin{align}
    \Delta w = \tilde{\alpha} \left(y^{(j)}-\hat{y}_w[x^{(j)}]\right) x^{(j)}.\label{eq:perceptron_discrete}
\end{align}
The \emph{learning rate} $\tilde{\alpha}>0$ determines the magnitude of the correction applied for each training example. The expression in parentheses can only take on the values $\{ 0, -1, 1\}$ with the zero corresponding to a correctly classified example and the non-zero values corresponding to the two different possible classification errors.

Usually there exist many separating hyperplanes for a given linear binary classification problem. The standard perceptron is only guaranteed to find one that works for the training set. It is possible to introduce a notion of optimality to this problem by considering the minimal distance (``margin'') of the training data to the found separating hyperplane. Maximization of this margin naturally leads to the ``support vector machine'' (SVM) algorithm \cite{Cortes1995SupportVector}. Although the SVM outperforms the perceptron in many classification tasks it does not lend itself to a hardware implementation as readily because it cannot be trained incrementally. It is this that makes the perceptron algorithm especially suited for a hardware implementation: We can convert the discrete update rule \eqref{eq:perceptron_discrete} to a differential equation
\begin{align}
    \dot{w}(t) = \alpha \left\{y(t)-\hat{y}_{w(t)}(t)\right\} x(t), \label{eq:perceptron_continuous}
\end{align}
and then construct a physical system that realizes these dynamics. 
In this continuous-time version the inputs are piece-wise constant $x(t) = x^{(j_t)},$  $y(t) = y^{(j_t)}$ and take on the same discrete values as above indexed by $j_t := \lceil \frac{t}{\Delta t} \rceil \in \{1,2,\dots, M = \frac{T}{\Delta t}\}.$

\subsection{The circuit modeling framework} 
\label{sub:the_model}
Circuits are fully described via Quantum Hardware Description Language (QHDL) \cite{Tezak2012Specification} based on Gough and James' SLH-framework \cite{Gough2009Series,Gough2008Quantum}.
To carry out numerical simulations for large scale networks, we derive a system of semi-classical Langevin equations based on the Wigner-transformation as described in \cite{Santori2014Quantum}. Note that there is a perfect one-to-one correspondence between nonlinear cavity models expressed via SLH and the Wigner method as long as the nonlinearities involve only oscillator degrees of freedom. There is ongoing research in our group to establish similar results for more general nonlinearities \cite{Hamerly2015FCM}.

Both the Wigner method and the more general SLH framework can be used to model networks of quantum systems where the interconnections are realized through bosonic quantum fields.
The SLH framework describes a system interacting with $n$ independent input fields in terms of a unitary scattering matrix $S$ parametrizing direct field scattering, a coupling vector $L=(L_1, L_2, \dots, L_n)^T$ parametrizing how external fields couple into the system and how the system variables couple to the output and a Hamilton operator inducing the internal dynamics. We summarize these objects in a triplet $(S, L, H).$ $L$ and $H$ are sufficient to parametrize any Schrödinger picture simulation of the quantum dynamics, e.g., the master equation for a mixed system state $\rho$ is given by
\begin{align}\label{eq:SLH_master}
  \dot \rho = - i[H, \rho] + \sum_{j=1}^n \left(L_j \rho L_j^\dagger - \frac12 \{L_j^\dagger L_j, \rho\}\right).
\end{align}
The scattering matrix $S$ is important when composing components into a network. In particular, the input-output relation in the SLH framework is given by
\begin{align}\label{eq:SLH_inout}
  dA_{\rm out} = S\, dA_{\rm in} + L\,dt,
\end{align} 
where the $dA_{\rm in/out,j},\, j=1,2,\dots, n$ are to be understood as quantum stochastic processes whose differentials can be manipulated via a quantum Ito calculus \cite{Gough2009Series}. 
The Wigner method provides a simplified, approximate description which is valid when all non-linear resonator modes are in strongly displaced states $\cite{Santori2014Quantum}.$
The simulations presented here were carried out exclusively at energy scales for which the Wigner method is valid, allowing us to scale to much larger system sizes than we could in a full SLH-based quantum simulation. This is because the computational complexity of the Wigner method scales at most quadratically (and in sparsely interconnected systems nearly linearly) with the number of components as opposed to the exponential state space scaling of a quantum mechanical Hilbert space. We nonetheless provide our models in both Wigner-method form and SLH form in anticipation that our component models will also be extremely useful in the full quantum regime.

In the Wigner-based formalism, a system is described in terms of time-dependent complex coherent amplitudes $\z(t)=(\zz{1}(t), \zz{2}(t),\dots, \zz{m}(t))^T$ for the internal cavity modes and external inputs $\bin(t) = (\binn{1}(t), \binn{2}(t), \dots, \binn{n}(t))^T$. These amplitudes relate to quantum mechanical expectations as $\langle \zz{j} \rangle \approx \langle a_j\rangle_{\rm QM},$ where $\langle \cdot \rangle$ denotes the expectation with respect to the Wigner quasi distribution and $\langle \cdot \rangle_{\rm QM}$ a quantum  mechanical expectation value. See \cite{Santori2014Quantum} for the corresponding relations of higher order moments.

To simplify the analysis, we exclusively work in a rotating frame with respect to all driving fields.
As in the SLH case we define output modes $\bout(t)$ that are algebraically related to the inputs and the internal modes. The full dynamics of the internal and external modes are then governed by a multi-dimensional Langevin equation 
\begin{align}
\dot{\z}(t) &= \left[\mathbf{A} \z(t) + \mathbf{a} + A_{\rm NL}(\z,t)\right]
	 + \mathbf{B} \bin(t), \label{eq:abnl}
\end{align}
as well as a purely algebraic, linear input-output relationship
\begin{align}
	\bout(t) &= \left[\mathbf{C} \z(t) + \mathbf{c}\right] + \mathbf{D}\bin(t) \label{eq:cd-io}.
\end{align}
The complex matrices $\mathbf{A}, \mathbf{B}, \mathbf{C}, \mathbf{D}$ as well as the constant bias input vectors $\mathbf{a}$ and $\mathbf{c}$ parametrize the linear dynamics, whereas the function $A_{\rm NL}(\z,t)$ gives the nonlinear contribution to the dynamics of the internal cavity modes.

Each input consists of a coherent, deterministic part and a stochastic contribution $\binn{j}(t)=\bincohh{j}(t) + \eta_j(t)$.
The stochastic terms $\eta_j(t) = \eta_{j,1}(t) + i \eta_{j, 2}(t)$ are assumed to be independent complex Gaussian white noise processes with correlation function $\langle \eta_{j,s}(t)\eta_{k,r}(t')\rangle = \frac{1}{4}\delta_{jk}\delta_{sr}\delta(t-t').$

The linearity of the input-output relationship in either framework \eqref{eq:SLH_inout} and \eqref{eq:cd-io} in the external degrees of freedom leads to algebraic rules for deriving reduced models for whole circuits of nonlinear optical resonators by concatenating component models and algebraically solving for their interconnections. \cite{Gough2008Quantum,Santori2014Quantum}
To see the basic component models used in this work see Appendix \ref{sec:component_models}. Netlists for composite components and the whole circuit will be made available at \cite{Tezak2014PerceptronFiles}.

\section{The Coherent Perceptron Circuit} 
\label{sec:the_all_optical_perceptron_circuit}

\begin{figure*}[hbt]
    \centering
        \includegraphics[width=12cm]{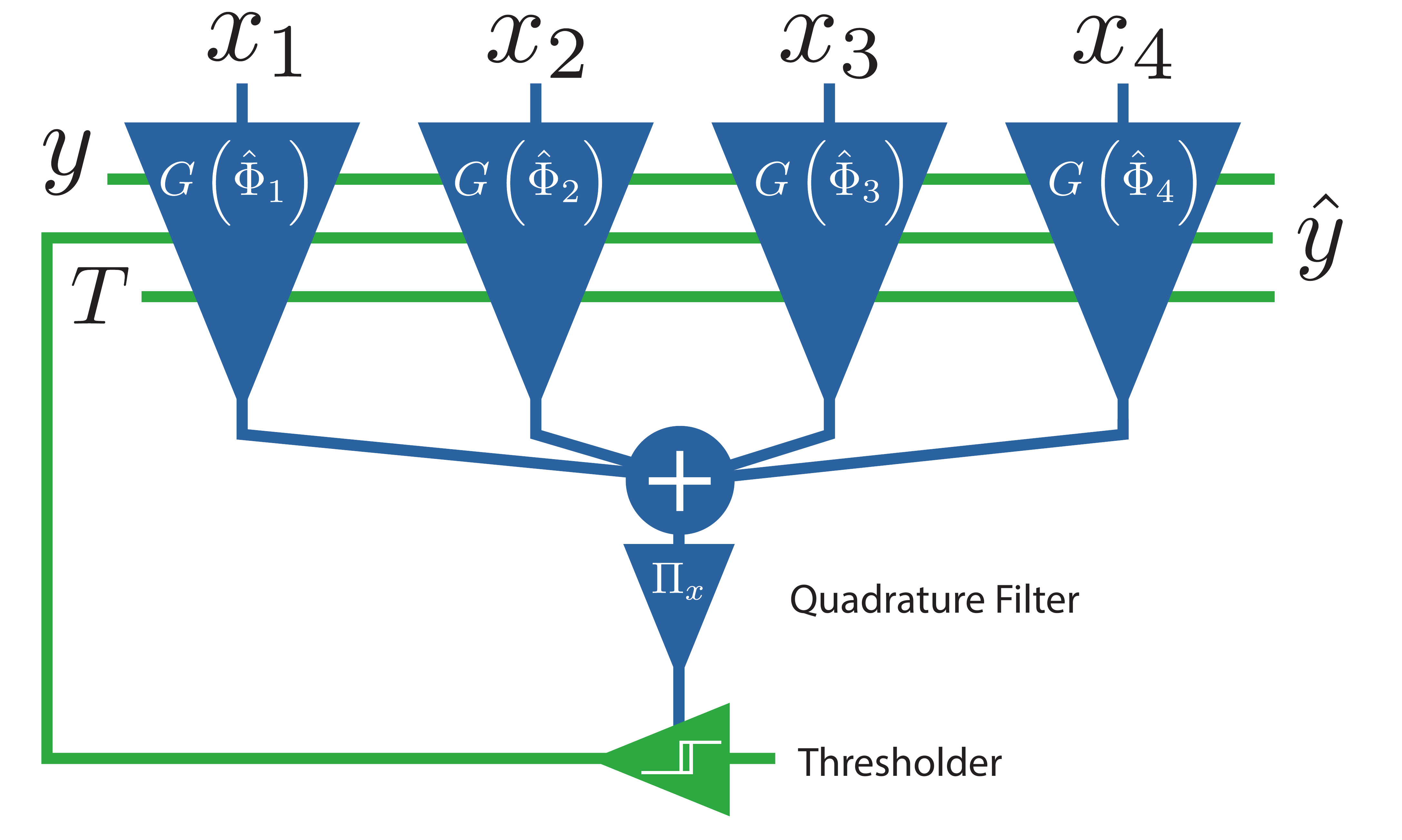}
    \caption{An example perceptron circuit consisting of $N=4$ programmable amplifiers for the coherent input vector $x = (x_1, x_2, x_3, x_4)^T$, a static mixing element that sums their output, a quadrature filter to remove the imaginary quadrature and a final thresholding element to generate the estimated binary class label $\hat{y}.$ The additional binary input $T$ controls whether the system is in \emph{training mode}, in which case the estimated class label $\hat{y}$ is compared to the true class label $Y$ which is provided as an additional input. When they differ, the programmable amplifiers receive a feedback signal to adjust their internal weights.}
    \label{fig:Perceptron}
\end{figure*}

The full perceptron's circuit is visualized in Figure~\ref{fig:Perceptron}. 
The input data $x$ to the perceptron circuit is encoded in the real quadrature of $N$ coherent optical inputs. 
Equation~\eqref{eq:perceptron_continuous} informs us what circuit elements are required for a hardware implementation by decomposing the necessary operations:
\begin{enumerate}
    \item Each input $x_j$ is multiplied by a weight $w_j$. 
    \item The weighted inputs are coherently added.
    \item The sum drives a thresholding element to generate the estimated class label $\hat{y}$.
    \item In the training phase (input $T=1$) the estimated class label $\hat{y}$ is compared with the true class label (input $Y$) and based on the outcome, feedback is applied to modify the weights $\{w_j\}$.
\end{enumerate}
The most crucial element for this circuit is the system that multiplies an input $x_j$ with a programmable weight $w_j$. This not only requires having a linear amplifier with tunable gain, but also a way to encode and store the continuous weights $w_j$.
In the following we outline one way how such systems can be constructed from basic nonlinear optical cavity models: Section~\ref{sec:variable_gain_amplifiers} presents an elegant way to construct a phase sensitive linear optical amplifier where the gain can be tuned by changing the amplitude of a bias input. 
In Section~\ref{sec:encoding_and_storing_the_gain} we propose using an above threshold non-degenerate optical parametric amplifier to store a continuous variable in the output phase of the signal (or idler) mode. In Section~\ref{sec:programmable_gain_amplifier} these systems are combined to realize an optical amplifier with \emph{programmable} gain, i.e., a control input can program its gain, which then stays constant even after the control has been turned off.
Finally, we present a simple model for all-optical switches based on a cavity with two modes that interact via a cross-Kerr-effect in Section~\ref{sec:optical_switches}. This element is used both for the feedback logic as well as the thresholding function to generate the class label $\hat{y}$.

\subsection{Tunable Gain Kerr-amplifier} 
\label{sec:variable_gain_amplifiers}

A single mode Kerr-nonlinear resonator driven by an appropriately detuned coherent drive $\epsilon$ can have a strongly nonlinear dependence of the intra-cavity energy on the drive power. When the drive of a single resonator is given by the sum of a constant large bias amplitude and a small signal $\epsilon=\frac{1}{\sqrt{2}} (\epsilon_0 +\delta\epsilon)$, the steady state reflected amplitude is $\epsilon'=\frac{1}{\sqrt{2}} (\eta\epsilon_0 + g_-(\epsilon_0) \delta\epsilon + g_+(\epsilon_0) \delta\epsilon^\ast) +O(\delta\epsilon^2)$, where $|\eta|\le 1$ with equality for the ideal case of negligible intrinsic cavity losses. The small signal thus experiences phase sensitive gain dependent on the bias amplitude and phase. We provide analytic expressions for the gain in Appendix \ref{ssub:single_mode_kerr}.

\begin{figure}[htb]
    \centering
        \subfigure[Amplifier circuit]{\label{fig:amp_circuit}\includegraphics[width=6cm]{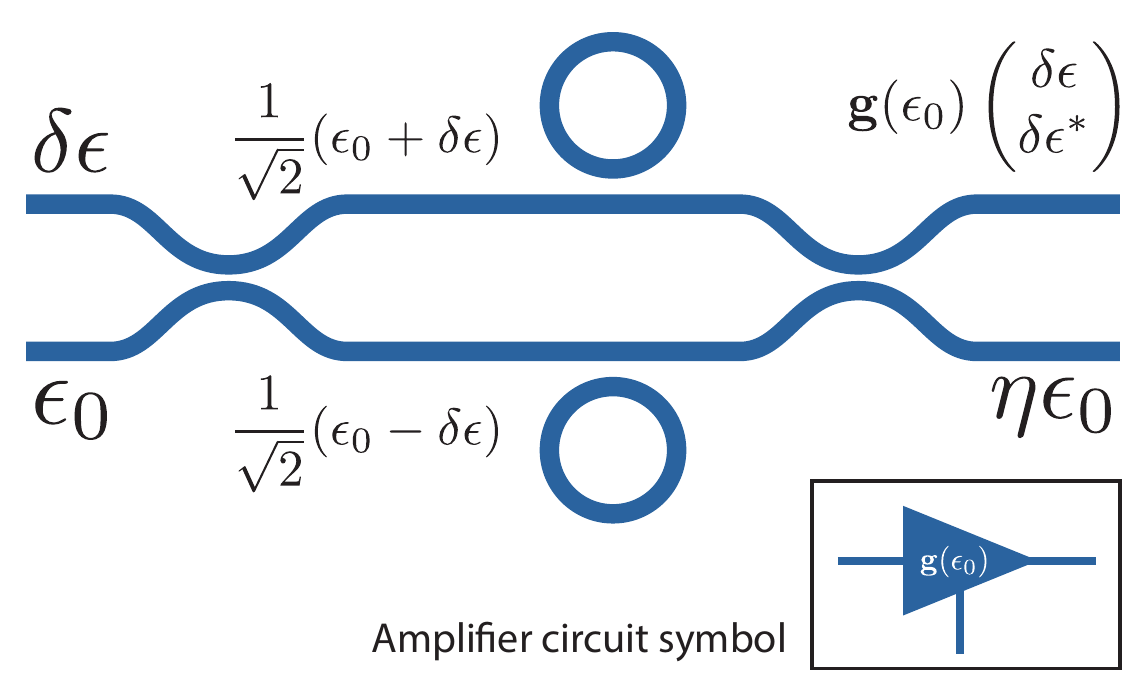}}
        \subfigure[Gain vs. bias]{\label{fig:amp_gain_bias}\includegraphics[width=6cm]{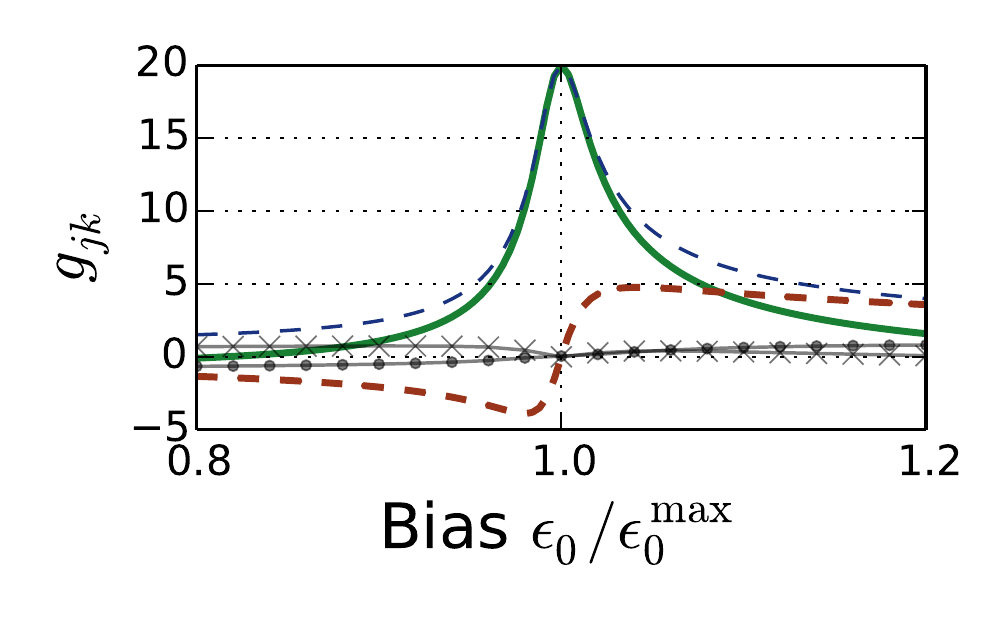}}
    \caption{(a) shows two identical single mode Kerr-nonlinear optical resonators symmetrically placed in the two arms of an interferometer. (b) gives the phase sensitive amplifier gain $g_{rr}(\epsilon_0)$ (green, solid) and the $g_{ir}(\epsilon_0)$ (red, dashed) as a function of the bias photon input rate normalized by the drive power at which dynamic resonance occurs. For completeness we also provide $g_{ri}$ (black X's) and $g_{ii}$ (black dots).
    The detuning has been chosen such that $g_{rr}^{\rm{max}} = g_{rr}(\epsilon_0^{\rm{max}})=20$. The dashed blue envelope gives the maximal input output gain achievable between any two signal quadratures at that bias. Note that $g_{rr}$ vanishes at $\epsilon_0/\epsilon_0^{\rm max}\approx 0.8$.}
    \label{fig:figures_Amplifier}
\end{figure}

Placing two identical resonators in the arms of an interferometer allows for isolating the signal and bias outputs even if their amplitudes vary by canceling the scattered bias in one output and the scattered signal in the other (cf.~Figure~\ref{fig:figures_Amplifier}). This highly symmetric construction, which generalizes to any other optical nonlinearity, ensures that the the signal output is linear in $\delta \epsilon$ up to third order\footnote{One can easily convince oneself that all even order contributions are scattered into the bias output.}.
If the system parameters are well-chosen, the amplifier gain depends very strongly on small variations of the bias amplitude. This allows to tune the gain from close to unity to its maximum value, which, for a given waveguide coupling $\kappa$ and Kerr coefficient $\chi$ depends on the drive detuning from cavity. For Kerr-nonlinear resonators there exists a critical detuning beyond which the system becomes bi-stable and exhibits hysteresis. This can be used for thresholding type behavior though as shown in \cite{Tait2013Dream} in this case it may be advantageous to reduce the symmetry of the circuit.
It is convenient to engineer the relative propagation phases such that at maximum gain, a real quadrature input signal $x\in \RR$ leads to an amplified output signal $x' = g_{rr}^{\rm{max}}x$ with no imaginary quadrature component (other than noise and higher order contributions). However, for different bias input amplitudes and consequently lower gain values the output will generally feature a linear imaginary quadrature component $x' = \left[g_{rr}(\epsilon_0) + i g_{ir}(\epsilon_0)\right]x$ as well. Figure \ref{fig:amp_gain_bias} demonstrates this for a particular choice of maximal gain.
We note that there exist previous proposals of using nonlinear resonator pairs inside interferometers to achieve desirable input-output behavior \cite{Tait2013Dream}, but to our knowledge, no one has proposed using these for signal/bias isolation and tunable gain.
To first order the linearized Kerr model is actually identical to a sub-threshold degenerate OPO model. This implies that it can be used to generate squeezed light and also that one could replace the Kerr-model by an OPO model.

An almost identical circuit, but featuring resonators with additional internal loss equal to the wave-guide coupling\footnote{In the photonics community this is referred to as \emph{critically coupled,} whereas the amplifier circuit would ideally be strongly \emph{overcoupled} such that additional internal losses are negligible.} and constantly biased to \emph{dynamic resonance} $\langle |\alpha|^2 \rangle_{\rm{ss}} = -\Delta/\chi$ can be used to realize a \emph{quadrature filter}, i.e., an element that has unity gain for the real quadrature and zero for the imaginary one. Now the quadrature filtered signal still has an imaginary component, but to linear order this only consists of transmitted noise from the additional internal loss.
While it would be possible to add one of these downstream of every tunable Kerr amplifier, in our specific application it is more efficient to add just a single one downstream of where the individual amplifier outputs are summed (cf. Section \ref{sec:thresholder}). This also reduces the total amount of additional noise introduced to the system.

\subsection{Encoding and Storing the Gain} 
\label{sec:encoding_and_storing_the_gain}
In the preceding section we have seen how to realize a tunable gain amplifier, but for programming and \emph{storing} this gain (or equivalently its bias amplitude) an additional component is needed. Although it is straightforward to design a multi-stable system capable of outputting a discrete set of different output powers to be used as the amplifier bias, such schemes would likely require multiple nonlinear resonators and it would be more cumbersome to drive transitions between the output states.

An alternative to such schemes is given by systems that have a continuous set of stable states. Recent analysis of continuous time recurrent neural network models trained for complex temporal information processing tasks has revealed multi-dimensional stable attractors in the internal network dynamics that are used to store information over time. \cite{Sussillo2013Opening}

A simple semi-classical nonlinear resonator model to exhibit this is given by a non-degenerate optical parametric oscillator (NOPO) pumped above threshold; for low pump input powers this system allows for parametric amplification of a weak coherent signal (or idler) input. In this case vacuum inputs for the signal and idler lead to outputs with zero expected photon number. Above a critical threshold pump power, however, the system down-converts pump photons into pairs of signal and idler photons. 

\begin{figure}[htb]
    \centering
    \subfigure[Combined bias]{\label{fig:phase_memory_bias}\includegraphics[width=5cm]{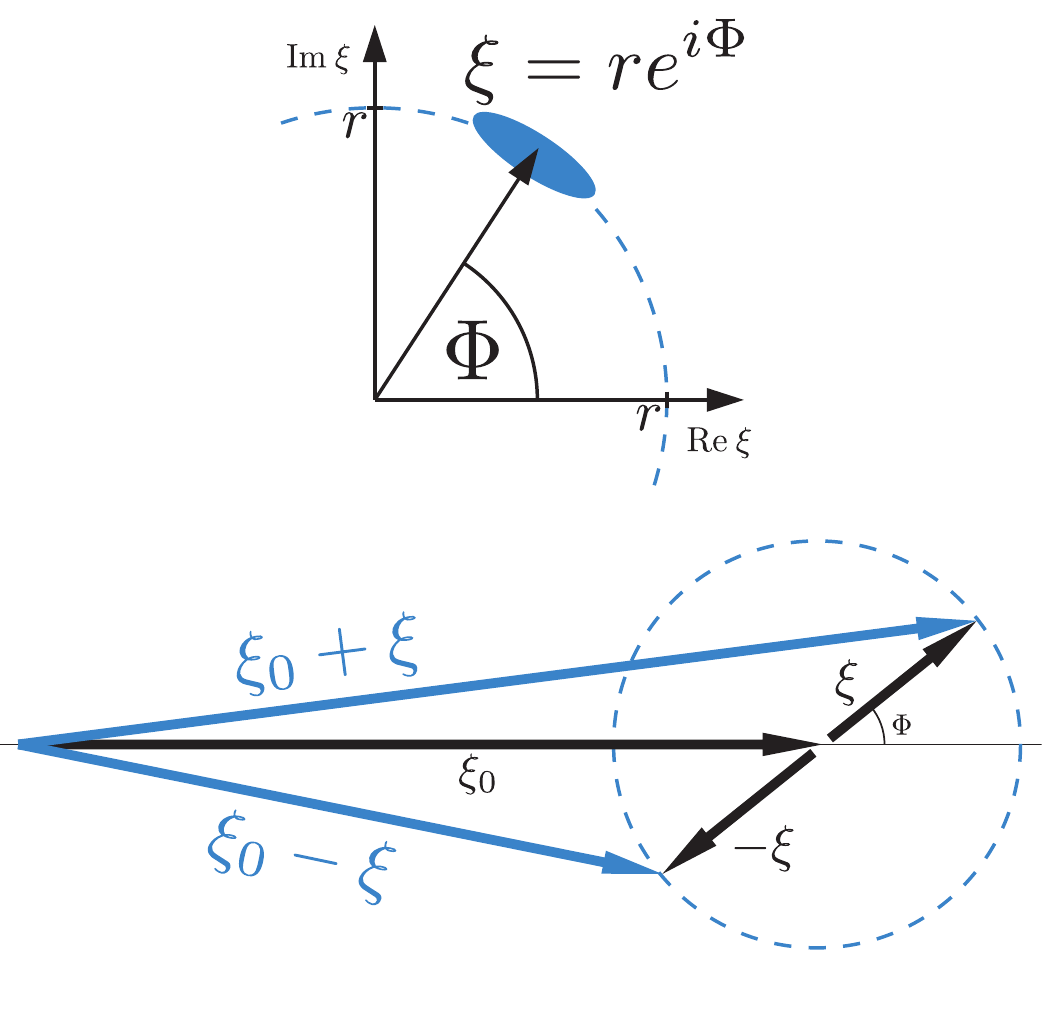}}
    \subfigure[Gain vs. OPO phase]{\label{fig:gain_vs_phase}\includegraphics[width=7cm]{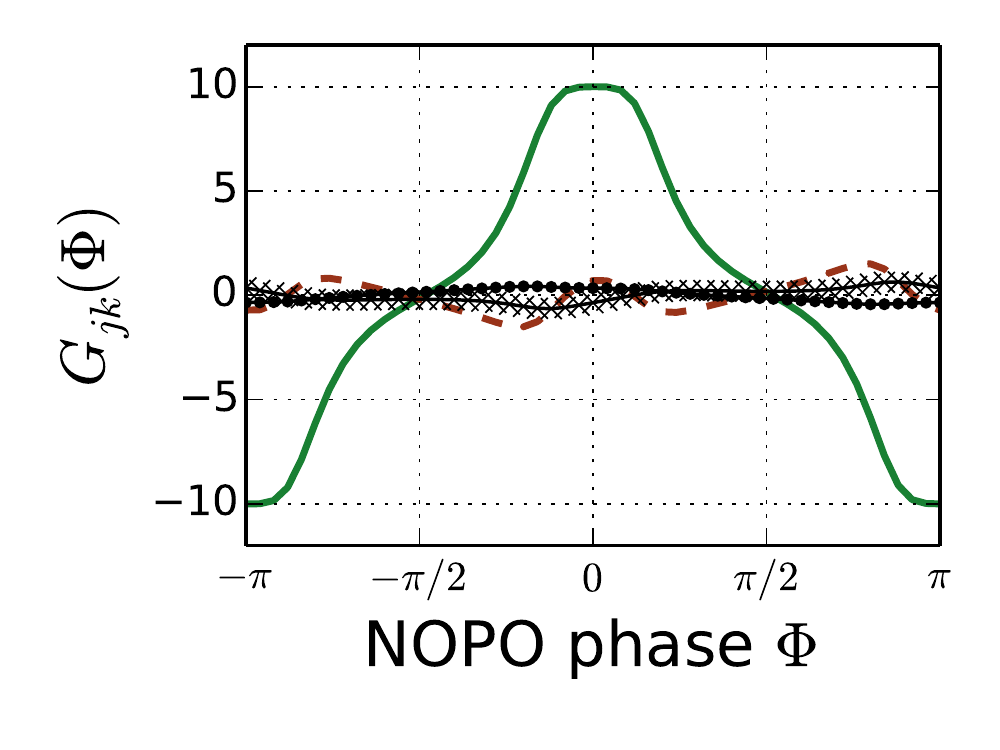}}    
    \caption{The NOPO's signal output $\xi=\sqrt{\kappa}\alpha_s$ lives on a circular manifold parametrized by $\Phi$ (a, upper figure). Vacuum input shot noise leads to small fluctuations perpendicular to the manifold and diffusion along it. Mixing this signal output with a constant bias offset on a beamsplitter produces two outputs with anti-correlated total amplitude (a, lower figure). When both outputs are used to drive a complementary pair of tunable amplifiers whose outputs are subtracted, the overall real-to-real quadrature gain (green) of the system varies from positive to negative values (b). We can also see that the real-to-imaginary gain (dashed red) stays small for all NOPO phases, which allows us to efficiently subtract it downstream by the quadrature filter. The imaginary to real and imaginary gains are also plotted.}
    \label{fig:figures_PhaseMemory}
\end{figure}

Due to an internal $U(1)$ symmetry of the underlying Hamiltonian (cf.~Appendix~\ref{ssub:nopo_model}), the signal and idler modes spontaneously select phases that are dependent on each other but independent of the pump phase. This implies that there exists a whole manifold of fix-points related to each other via the symmetry transformation $(\alpha_s, \alpha_i)\to(\alpha_s e^{i\phi}, \alpha_i e^{-i\phi})$, where $\alpha_s$ and $\alpha_i$ are the rotating frame signal and idler mode amplitudes, respectively.
Consequently the signal output of an above threshold NOPO lives on a circular manifold (cf Figure~\ref{fig:figures_PhaseMemory}).

Vacuum shot noise on the inputs leads to phase diffusion with a rate of $\gamma_\Phi = \frac{\kappa}{8n_0}$, where $\kappa$ is the signal and idler line width and $n_0$ is the steady state intra cavity photon number in either mode. We point out that this diffusion rate does not directly depend on the strength of the nonlinearity which only determines how strongly the system must be pumped to achieve a given intra cavity photon number $n_0$.

A weak external signal input breaks the symmetry and biases the signal output phase towards the external signal's phase. This allows for changing the programmed phase value.

Finally, we note that parametric oscillators can also be realized in materials with vanishing $\chi_2$ nonlinearity. They have been successfully realized via four-wave mixing (i.e., exploiting a $\chi_3$ nonlinearity) in \cite{Kippenberg2004KerrNonlinearity,Savchenkov2004Low,Haye2007Optical} and even in opto-mechanical systems \cite{Cohen2014Phonon} in which case the idler mode is given by a mechanical degree of freedom.

In principle any nonlinear optical system that has a stable limit cycle could be used to store and encode a continuous value in its oscillation phase. Non-degenerate parametric oscillators stand out because of their theoretical simplicity allowing for a `static' analysis inside a rotating frame.

\subsection{Programmable Gain Amplifier} 
\label{sec:programmable_gain_amplifier}
Combining the circuits described in the preceding sections allows us to construct a fully programmable phase sensitive amplifier. In Figure \ref{fig:amp_gain_bias} we see that there exists a particular bias amplitude at which the real to real quadrature gain vanishes $g_{rr}(\epsilon_{0}^{\rm{min}}) = 0$. We combine the NOPO signal output $\xi=r e^{i\Phi}$ with a constant phase bias input $\xi_0$ (cf. Figure \ref{fig:phase_memory_bias}) on a beamsplitter such that the outputs vary between zero gain and the maximal gain bias values $\left|\frac{\xi_0 \pm r e^{i\Phi}}{\sqrt{2}}\right| \in [\epsilon_{0}^{\rm{min}}, \epsilon_{0}^{\rm{max}}]$. To realize both positive and negative gain, we use the second output of that beamsplitter to bias another tunable amplifier. The two amplifiers are always biased oppositely meaning that one will have maximal gain when the other's gain vanishes and vice versa.
The overall input signal is split and sent through both amplifiers and then re-combined with a relative $\pi$ phase shift. This complementary setup leads to an overall effective gain tunable within $G_{rr}(\Phi) \in [-\frac{g_{rr}^{\rm max}}{2}, \frac{g_{rr}^{\rm max}}{2}]$ (cf.~ Figure \ref{fig:gain_vs_phase}).

\begin{figure}[htbp]
    \centering
        \includegraphics[width=10cm]{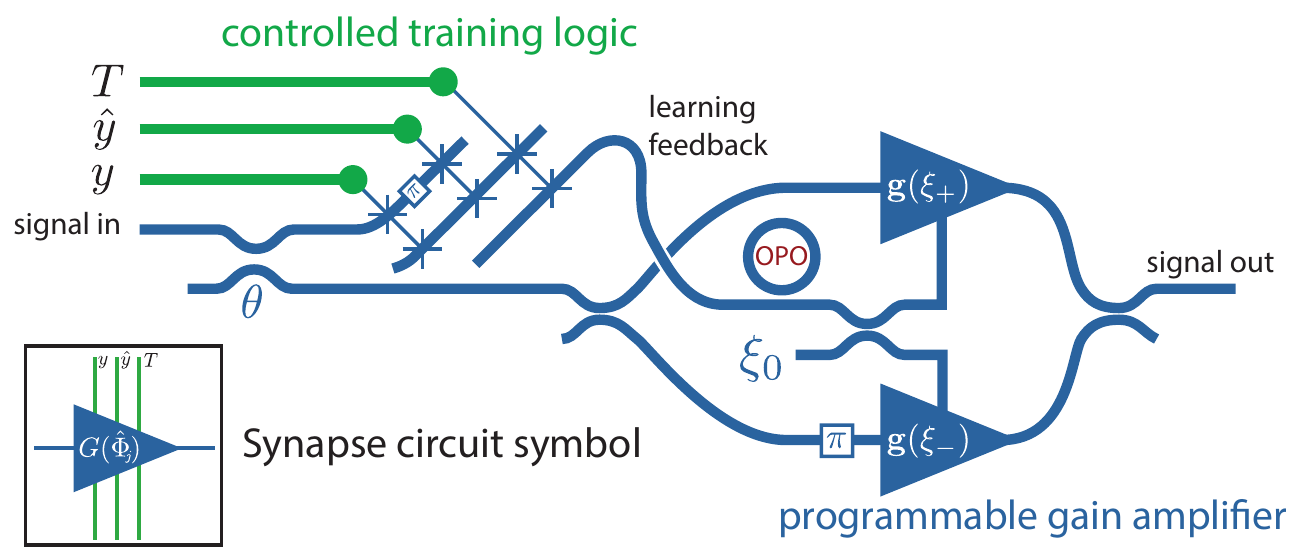}
    \caption{Synapse circuit composed of a programmable amplifier and feedback logic (cf. Section \ref{sec:optical_switches}) that implements the perceptron learning feedback \eqref{eq:perceptron_continuous} for a single weight. The upper amplifier when biased optimally leads to positive gain whereas the lower amplifier leads to negative gain due to the additional $\pi$ phase shift.}
    \label{fig:synapse}
\end{figure}

In Figure~\ref{fig:synapse} we present both the complementary pair of amplifiers and the NOPO used for storing the bias as well as some logic elements (described in Section \ref{sec:optical_switches}) used for implementing conditional training feedback. We call the full circuit a synapse because it features programmable gain and implements the perceptron's conditional weight update rule.

The resulting synapse model is quite complex and certainly not optimized for a minimal component number but rather the ease of theoretical analysis. A more resource efficient programmable amplifier could easily be implemented using just two or three nonlinear resonators. E.g., inspecting the the real to imaginary quadrature gain $g_{ir}(\epsilon_0)$ in Figure \ref{fig:amp_gain_bias} we see that close to $\epsilon_0^{\rm{max}}$ it passes through zero fairly linearly and with an almost symmetric range. This indicates that we could use a single tunable amplifier to realize both positive and negative gain. Using only a single resonator for the tunable amplifier could work as well, but it would require careful interferometric bias cancellation and more tedious upfront analysis.             
We do not think it is feasible to use just a single resonator for both the parametric oscillator and the amplifier because any amplified input signal would have an undesirable back-action on the oscillator phase.

\subsection{Optical Switches} 
\label{sec:optical_switches}
The feedback to the perceptron weights (cf. Equation \eqref{eq:perceptron_continuous}) is conditional on the binary values of the given and estimated class labels $y$ and $\hat{y}$, respectively. The logic necessary for implementing this can be realized by means of all-optical switches. There have been various proposals and demonstrations \cite{Poustie2000Demonstration,Nozaki2010SubFemtojoule} of all-optical gates/switches and quantum optical switches \cite{Milburn1989Quantum}.

\begin{figure}[htb]
    \centering
        \subfigure[Fredkin gate and thresholder]{\label{fig:fredkin_schem}\includegraphics[width=6cm]{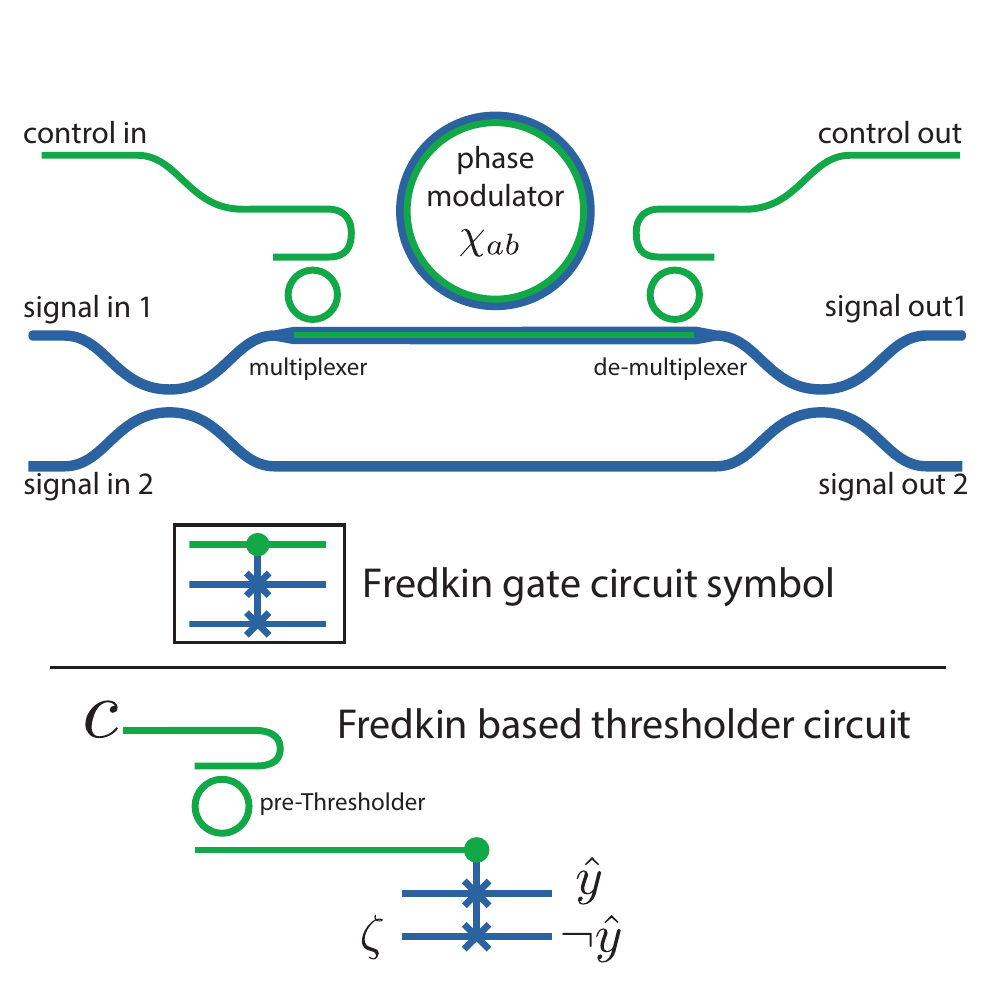}}
        \subfigure[Thresholder input/output]{\label{fig:fredkin_thresh}\includegraphics[width=6cm]{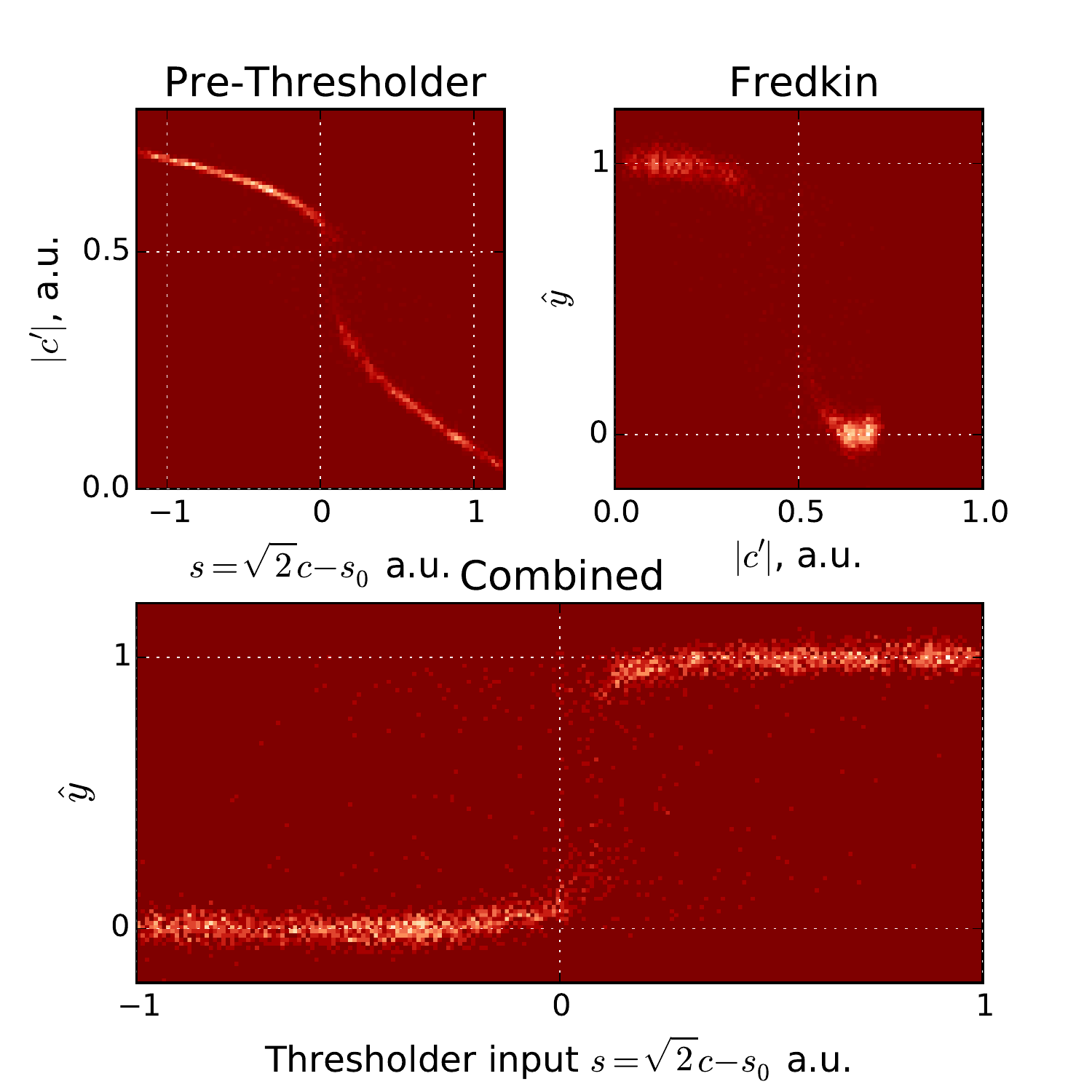}}
    \caption{In the upper graphic of (a) we present a schematic for Fredkin gate based on a two mode cross-Kerr-nonlinear resonator. The lower graphic shows how this circuit can be pre-pended with a single mode nonlinear resonator to better approximate a thresholding response. In (b) we present the input output characteristic of the prepended resonator (upper left), the Fredkin gate (upper right) and the combined input output relationship between the inner product amplitude $s$ and the estimated state label $\hat{y}$.}
    \label{fig:figures_Fredkin}
\end{figure}

The model that we assume here (cf. Figure \ref{fig:figures_Fredkin}) is to use two different modes of a resonator that interact via a cross-Kerr-effect, i.e., power in the control mode leads to a refractive index shift (or detuning) for the signal mode. The index shift translates to a control mode dependent phase shift of a scattered signal field yielding a controlled optical phase modulator. Wrapping this phase modulator in a Mach-Zehnder interferometer then realizes a controlled switch: If the control mode input is in one of two different states $|\xi| \in {0, \xi_0}$, the signal inputs are either passed through or switched. This operation is often referred to as a \emph{controlled swap} or Fredkin gate \cite{Fredkin1982Conservative} which was originally proposed for realizing reversible computation. This dispersive model has the advantage that the control input signal can be reused.

Note that at control input amplitudes significantly different from the two control levels the outputs are coherent mixtures of the inputs, i.e., the switch then realizes a tunable beamsplitter.

Finally, we point out that using two different (frequency non-degenerate) resonator modes has the advantage that the interaction between control and signal inputs is phase insensitive which greatly simplifies the design and analysis of cascaded networks of such switches.


\subsection{Generation of the Estimated Label} 
\label{sec:thresholder}

The estimated classifier label $\hat{y}$ should be a step function applied to the inner product of the weight vector and the input. In the preceding sections we have shown how individual inputs $x_j$ can be amplified with programmable gain to give $\tilde{s}_j = \tilde{G}(\Phi_j)x_j$, thus realizing the individual contributions to the inner product. 
These are then summed on an $n$-port beamsplitter that has an output which gives the uniformly weighted sum $\tilde{s} := \frac{1}{\sqrt{N}}\sum_{k=1}^N \tilde{G}(\Phi_k)x_k$.

The gain factors $\tilde{G}(\Phi_k) = G_{rr}(\Phi_k) + i G_{ir}(\Phi_k)$ generally have an unwanted imaginary part which we subtract by passing the summed output through a \emph{quadrature filter} circuit (cf. the last paragraph of Section \ref{sec:variable_gain_amplifiers}), which has unit gain for the real quadrature and zero gain for the imaginary quadrature leading to an overall output $s = {\rm Re} \, \tilde{s} = \frac{1}{\sqrt{N}}\sum_{k=1}^N G_{rr}(\Phi_k)x_k$. The thresholding circuit should now produce a high output if $s>0$ and a zero output if $s \le 0$.

It turns out that the optical Fredkin gate described in the previous section already works almost as a two mode thresholder, where the control input leads to a step-like response in the signal outputs: 
A constant signal input amplitude which encodes the logical `1' state is applied to one of the signal inputs. When the control input amplitude is varied from zero to $\xi_0$, the signal output turns on fairly abruptly at some threshold $\xi_{\rm th} < \xi_0$. 
To make the thresholding phase sensitive, the control input is given by the sum of $s$ and a constant offset $s_0$ that provides a phase reference: $c = \frac{1}{\sqrt{2}}(s + s_0)$.

For a Fredkin gate operated with continuous control inputs the signal output is almost zero for a considerable range of small control inputs. However, for very high control inputs, i.e., significantly above $\xi_0$, the signal output decreases instead of staying constant as would be desirable for a step-function like profile.
We found that this issue can be addressed by transmitting the control input through a single mode Kerr-nonlinear cavity, with resonance frequency chosen such that the transmission gain $|c'/c|$ is peaked close to $c'=\xi_0$. For input amplitudes larger than $c$, the transmission gain is lower (although $|c'|$ still grows monotonically with $|c|$) which extends the input range over which the subsequent Fredkin gate stays in the on-state.

\section{Results} 
\label{sec:results}
\begin{figure}[htbp]
  \centering
  \subfigure[Error rate vs. hardness]{\label{fig:error_rate_GDA}\includegraphics[width=0.45\textwidth]{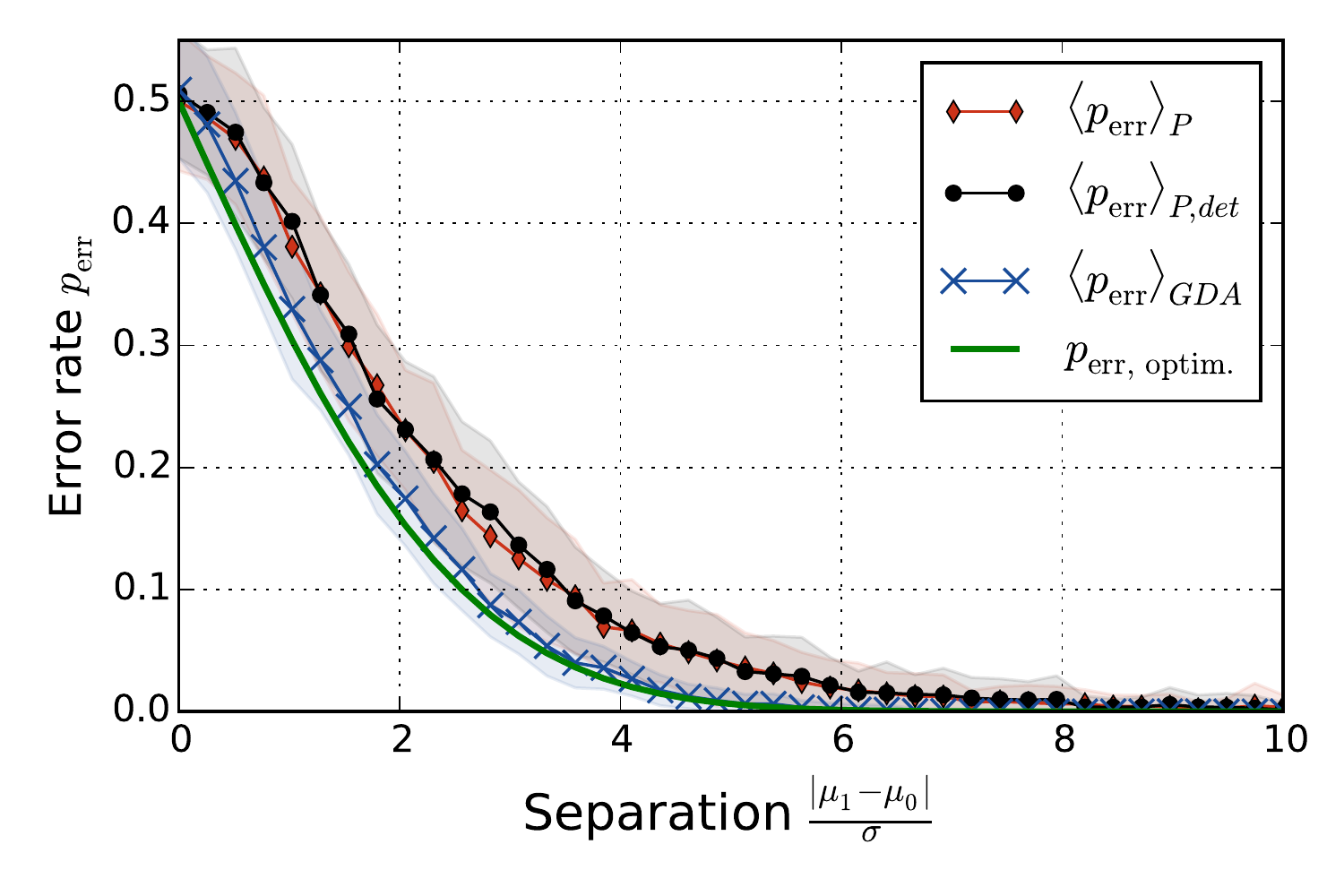}}
  \subfigure[Error rate vs. learning paramaters]{\label{fig:error_rate_learning_rate}\includegraphics[width=0.45\textwidth]{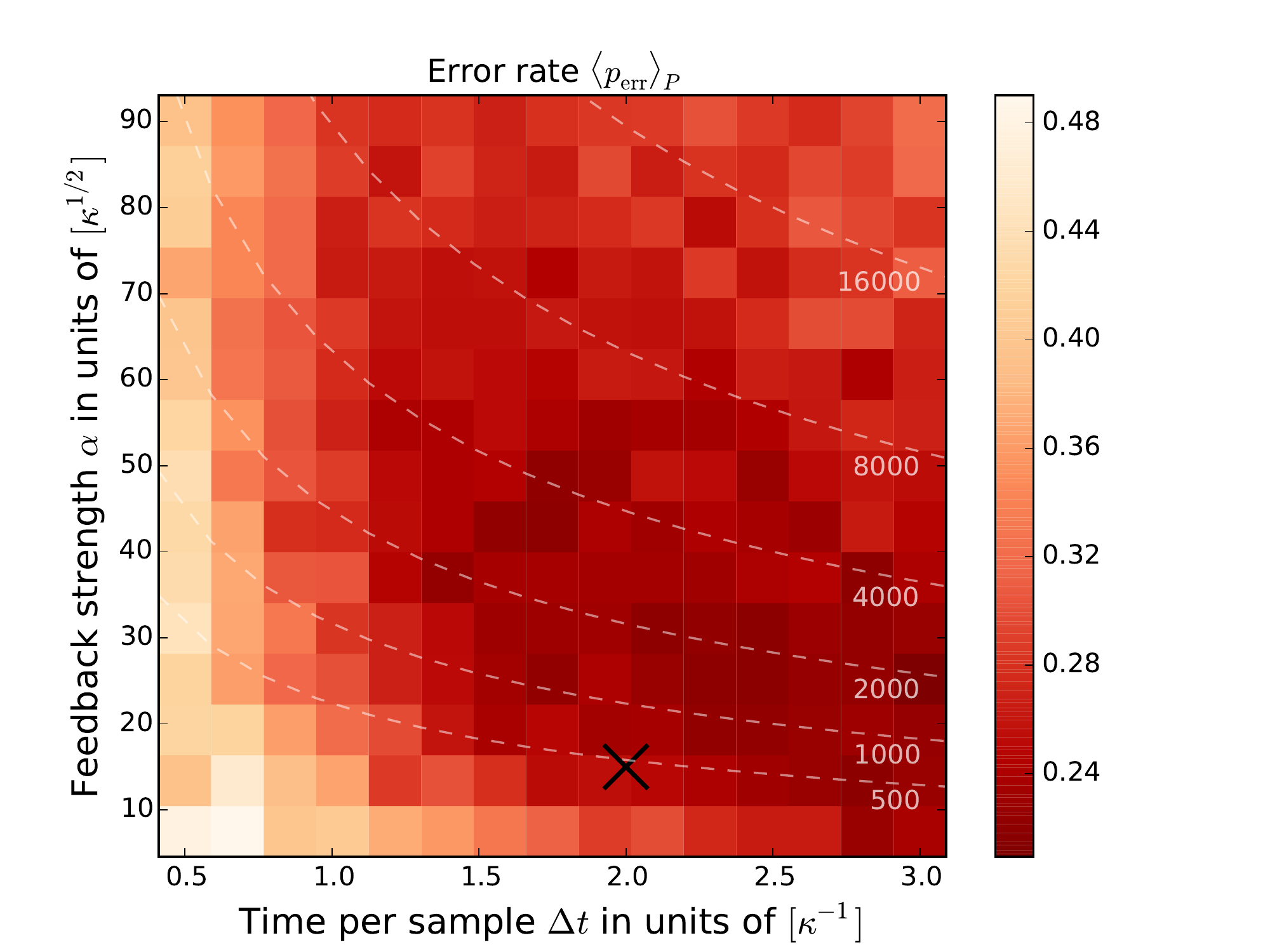}}
  \caption{The perceptron's error rate vs the difficulty of the classification task and as a function of the parameters determining the learning rate. In Figure (a) we compare the unoptimized performance of the perceptron circuit (red diamonds) to the optimal performance bound (solid, green) as well as a GDA (blue X's) trained on the same number of training examples. We show averages over $100$ trials at each cluster separation. The GDA data was similarly averaged over $100$ trials. The transparent envelopes indicate the sample standard deviation. The black dots show the perceptron performance when simulated without shot noise. We see that the shot noise has very little effect. In Figure (b) we plot the average error rate (averaged over $50$ trials) at fixed cluster separation $\| \mu_1 - \mu_0 \|_2 / \sigma = 2$ for various values of the time interval $\Delta t$ for which each data sample is presented to the circuit as well as the strength of the training feedback $\alpha$. The total number of feedback photons $N_{fb} = |\alpha|^2 \Delta t$ per sample is constant along the faint dashed lines and the actual value is indicated on the right. A good choice of parameters is characterized both by low feedback power (small $|\alpha|^2$) and high input rate (low sample time $\Delta t$) while still resulting in a low classification error rate. The X marks the parameters used for the results in (a) and the previous Figures.}
  \label{fig:error_rate_gda_opt}
\end{figure}

The perceptron's SDEs where simulated using a newly developed custom software package named QHDLJ \cite{Tezak2014Qhdlj} implemented in Julia \cite{Bezanson2014Julia} which allows allows for dynamic compilation of circuit models to LLVM \cite{Lattner2004Llvm} bytecode that runs at speed comparable to C/C++. 
All individual simulations can be carried out on a laptop, but the results in Figure~\ref{fig:error_rate_gda_opt} were obtained by averaging over the results of 100 stochastic simulation run on an HP ProLiant server with 80 cores. The current version of QHDLJ uses one process per trajectory, but the code could easily be vectorized.

\begin{figure}[htbp]
  \centering
  \includegraphics[width=0.95\textwidth]{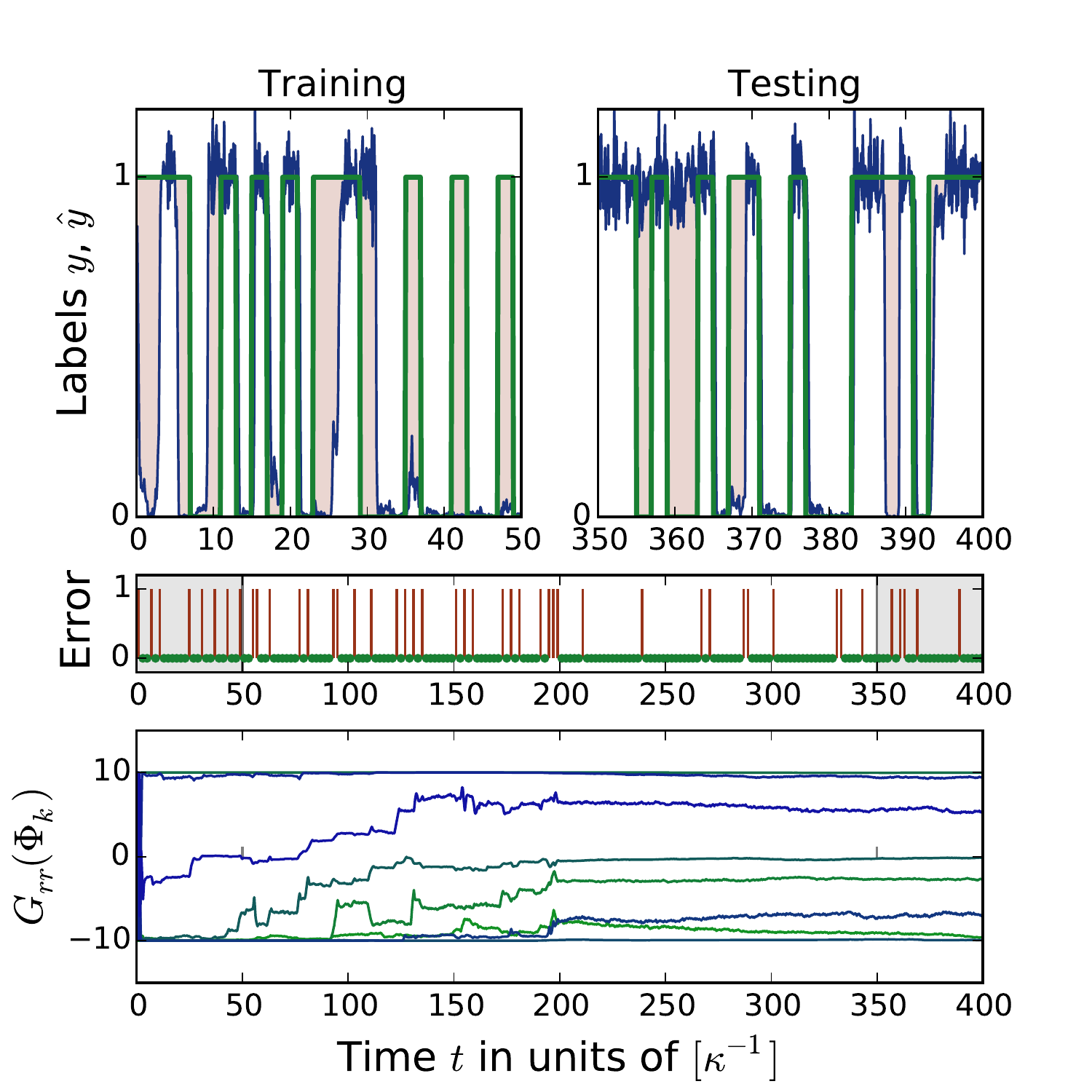}
  \caption{Single trajectory divided into a training interval $0 \le t \le M_{\rm train}\Delta t$ during which the learning feedback is active and a test interval $M_{\rm train}\Delta t <  t \le M_{\rm test}\Delta t$. During training and testing, respectively, the system is driven by $M_{\rm train} = M_{\rm test} = 100$ separate input states which are held constant for an interval $\Delta t = 2 \kappa^{-1}$. The estimated class label is discretized by averaging the output intensity over each input interval, dividing the result by the intensity $|\zeta|^2$ corresponding to the logical `1' output state and rounding.
  The upper panel compares the correct class label $y$ (green) with the estimated class label $\hat{y}$ (black) during training and testing, respectively. The area between them indicates errors or at least lag of the estimator and is shaded in light red. The second panel shows occurrences of classification errors (red vertical bars). The slight shading near the beginning and the end of the trajectory in the second panel visualizes the segments corresponding to the upper left and right panel, respectively. The third panel shows the learned linear amplitude gains for each synapse. After the learning feedback is turned off at $t=M_{\rm train}\Delta t$, they diffuse slightly due to optical shot noise.}
  \label{fig:classification_trajectory}
\end{figure}

In Figure \ref{fig:classification_trajectory} we present an example of a single application of an $N=8$ perceptron including both a learning stage with pre-labeled training data and a classification testing stage in which the perceptron's estimated class labels are compared with their correct values.
The data to be classified here are sampled from a different $8-$ dimensional Gaussian distribution for each class label with their mean vectors separated by a distance $ \| \mu_1 - \mu_0 \|_2 / \sigma = 2$ relative to the standard deviation of both individual clusters.
For each sample the input was held constant for a duration $\Delta t = 2 \kappa^{-1}$ where $\kappa$ is the NOPO signal and idler line width. The perceptron was first trained with $M_{\rm train}=100$ training examples and subsequently tested on $M_{\rm test}=100$ test examples with the learning feedback turned off.

\begin{figure}[htbp]
  \centering
  \includegraphics[width=\textwidth]{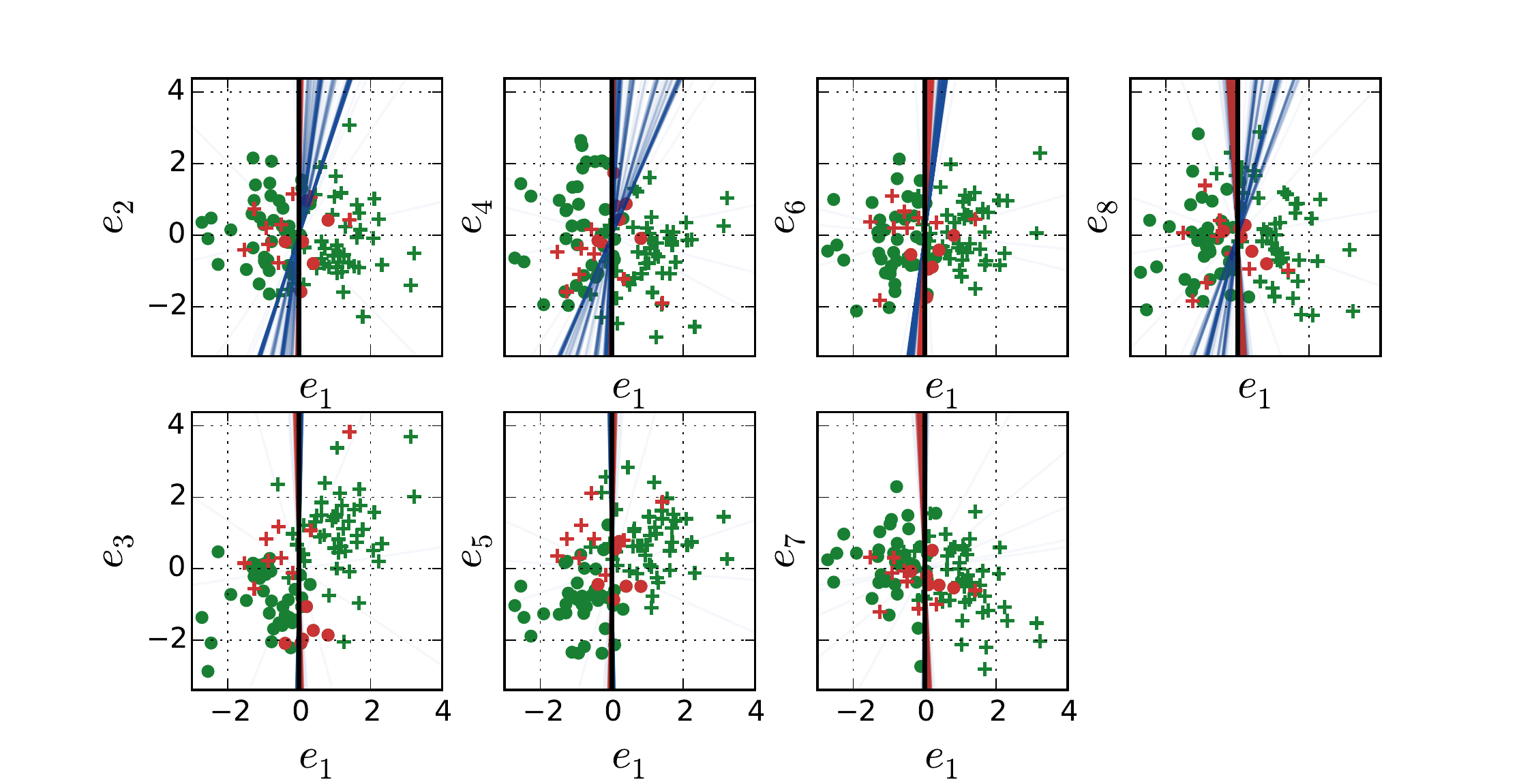}
  \caption{Projection of training data and classification boundaries. The data has been rotated such that the $s_1$ coordinate lines up with the learned normal vector of the separating hyperplane. Incorrectly classified data are plotted in red. The faint blue (red) lines visualize the evolution of the classifier boundary during training (testing).}
  \label{fig:classification_boundaries}
\end{figure}

In Figure \ref{fig:classification_boundaries} we visualize linear projections of the testing data as well as the estimated classification boundaries. We can see that the classifier performs very well far away from the decision boundary. Close to the decision boundary there are some misclassified examples. 
We proceed to compare the performance of the classifier to the theoretically optimal performance achievable by any classifier and with the optimal classifier for this scenario, Gaussian Discriminant Analysis (GDA) \cite{Fisher1936Use,Mclachlan1992Discriminant}, implemented in software. Using the identical perceptron model as above and an identical training/testing procedure, we estimate the error rate $p_{\rm err} = \mathbb{P}[y\ne \hat{y}]$ of the trained perceptron as a function of the cluster separation $ \| \mu_1 - \mu_0 \|_2 / \sigma$. The results are presented in Figure~\ref{fig:error_rate_GDA}. 
Identically distributed training and testing data was used to evaluate the performance of the GDA algorithm and both results are compared to the theoretically optimal error rate for this discrimination task, which can be computed analytically to be $p_{\rm err,\, optim.} = \frac{1}{2}{\rm erfc}\left(\frac{\| \mu_1 - \mu_0 \|_2}{\sqrt{8}\sigma}\right),$ where ${\rm erfc}(x) = \frac{2}{\sqrt{\pi}} \int_x^\infty e^{-u^2} {du}$ is the complementary error function.
We see that the all-optical perceptron's performance is comparable to GDA's performance for this problem and both algorithms attain performance close to the theoretical optimum.

The learning rate of the perceptron is determined by two things, the overall strength of the learning feedback as well as the time for which each example is presented to the circuit. In Figure \ref{fig:error_rate_learning_rate} we plot the estimated error rate for varying feedback strength and duration. As can be expected intuitively, we find that there are trade-offs between speed (smaller $\Delta t$ preferable) and energy consumption (smaller $\alpha$ preferable).
\subsection{Time scales and power budget} 
\label{sec:power_budget_and_energy_scalings}
Here we roughly estimate the power consumption of the whole device and discuss how to scale it up to a higher input dimension.

Any real-world implementation will depend strongly on the engineering paradigm, i.e., the choice of material/nonlinearity as well as the engineering precision, but based on recently achieved progress in nonlinear optics we will estimate an order of magnitude range for the input power.

The signal and feedback input power to the circuit will scale linearly in the number of synapses $N$. 

The bias inputs for the amplifiers has to be larger than the signal to ensure linearly operation, but it should be expected that some of the scattered bias amplitudes can be reused to power multiple synapses.

In our models we have defined all rates relative to the line width of the signal and idler mode of the NOPO, because this is the component that should necessarily have the smallest decay rate to ensure a long lifetime for the memory.

All other resonators are employed as nonlinear input-output transformation devices and therefore a high bandwidth (corresponding to much lower loaded quality factor) is necessary for achieving a high bit rate. For our simulations we typically assumed quality factors that were lower than the NOPO's by 1-2 orders of magnitude.

Based on self-oscillation threshold powers reported in \cite{Kippenberg2004KerrNonlinearity,Haye2007Optical,Levy2009CmosCompatible,Razzari2009CmosCompatible} and the switching powers of \cite{Nozaki2010SubFemtojoule} we estimate the necessary power per synapse to be in the range of $\sim 10-100\mu{\rm Watt}$. By re-using the scattered pump and bias fields it should be possible to reduce the power consumption per amplifier even further.
Even for the continuous wave signal paradigm we have assumed (as opposed to pulsed/spiking signals such as considered in \cite{Vaerenbergh2012Cascadable}) the devices proposed here could be competitive with the current state of the art CMOS-based neuromorphic electrical circuits \cite{Cassidy2014RealTime}.

In the simulations for the $8-$dimensional perceptron our input rate for training data was set to $\Delta t^{-1} = \frac{\kappa}{2}$. This value corresponds to roughly ten times the average feedback delay time between arrival of an input pattern and the conditional switching of the feedback logic upon arrival of the generated estimated state label $\hat{y}$. This time can be estimated as $\tau_{fb}(n) \approx G_{\rm max}\kappa_A^{-1} + \kappa_{QF}^{-1} + \kappa_{\rm thresh}^{-1} + n \kappa_F^{-1}$, where $n$ is the index of the synaptic weight, $G_{\rm max}$ is the amplifier gain range and $\kappa_A, \kappa_{QF}, \kappa_{\rm thresh}$ and $\kappa_F$ are the line widths of the amplifier, quadrature filter, the combined thresholding circuit (cf.~Figure~\ref{fig:figures_Fredkin}) and the feedback Fredkin gates. There is a contribution scaling with $n$ because the feedback traverses the individual weights sequentially to save power. 

When scaling up the perceptron to a higher dimension while retaining approximately the same input signal powers, it is intuitively clear that the combined `inner product' signal amplitude $s$ scales as $s\propto \sqrt{N}s_1$, where $s_1$ is the signal amplitude for a single input. This allows to similarly scale up the amplitude $\zeta_0$ of the signal encoding the generated estimated state label $\hat{y}$ and consequently the bandwidth of the feedback Fredkin gates that it drives. A detailed analysis reveals that the Fredkin gate threshold scales as $\sqrt{N}$, in particular we find that $ \sqrt{|\chi|}\zeta_0  \propto \kappa_F \propto \sqrt{|\chi|}\xi_0  \propto \kappa_{\rm thresh}\propto \sqrt{|\chi|} s\propto \sqrt{N|\chi|}s_1$. 
The first two scaling relationships are due to the constraints on the Fredkin gate construction (cf.~Appendix \ref{ssub:two_mode_kerr}), the next two scaling relationships follow from demanding that the additional thresholding resonator be approximately dynamically resonant at the highest input level (cf.~Appendices \ref{ssub:single_mode_kerr} and \ref{ssub:two_mode_kerr}). The last proportionality is simply due to the amplitude summation at the $N$-port beamsplitter.

This reveals that when increasing $N$ the perceptron as constructed here would have to be driven at a lower input bit rate scaling as $\Delta t^{-1} \propto N^{-\frac12}$ or alternatively be driven with higher signal input powers. 
A possible solution that could greatly reduce the difference in arrival time $\sim \kappa_F^{-1}$ at each synapse could be to increase the waveguide-coupling to the control signal and thus decrease the delay per synapse. 
The resulting increase in the required control amplitude $\zeta_0$ can be counter-acted with feedback, i.e., by effectively creating a large cavity around the control loop. 
When even this strategy fails one could add fan-out stages for $\hat{y}$ which introduce a delay that grows only logarithmically with $N$.

Finally, we note that the bias power of all the Kerr-effect based models considered here scales inversely with the respective nonlinear coefficient $\{|\zeta_0|^2, |s|^2\} \times |\chi| \sim {\rm const}$ when keeping the bandwidth fixed. This implies that improvements in the non-linear coefficient translate to lower power requirements or alternatively a faster speed of operation.

\section{Conclusion and Outlook} 
\label{sec:conclusion_and_outlook}

In conclusion we have shown how to design an all-optical device that is capable of supervised learning from input data, by describing how tunable gain amplifiers with signal/bias isolation can be constructed from nonlinear resonators and subsequently combined with self-oscillating resonators to encode the programmed amplifier gain in their oscillation phase. By considering a few additional nonlinear devices for thresholding and all-optical switching we then show how to construct a perceptron, including the perceptron feedback rule. To our knowledge this is the first end-to-end description of an all-optical circuit capable of learning from data.
We have furthermore demonstrated that despite optical shot-noise it nearly attains the performance of the optimal software algorithm for the classification task that we considered. Finally, we have discussed the relevant time-scales and pointed out how to scale the circuit up to large input dimensions while retaining the signal processing bandwidth and a low power consumption per input.

Possible applications of an all-optical perceptron are as the trainable output filter of an optical reservoir computer or as a building block in a multi-layer all-optical neural network.
 
The programmable amplifier could be used as a building block to construct other learning models that rely on continuously tunable gain such as Boltzmann machines and hardware implementations of message passing algorithms.

An interesting next step would be to design a perceptron that can handle inputs at different carrier frequencies. In this case wavelength division multiplexing (WDM) might allow to significantly reduce the physical footprint of the device.

A simple modification of the perceptron circuit could autonomously learn to invert linear transformations that were applied to its input signals. This could be used for implementing a circuit capable of solving linear regression problems. In combination with a multi-mode optical fibers such a device could also have applications for all-optical sensing. 

Finally, an extremely interesting question is whether harnessing quantum dynamics could lead to a performance increase. We hope to address these ideas in future work.


\section*{Competing interests}
  The authors declare that they have no competing interests.

\section*{Acknowledgements}

\label{sub:acknowledgements}
This work is supported by DARPA-MTO under award no. N66001-11-1-4106. N.T. acknowledges support from a Stanford Graduate Fellowship. We would also like to thank Ryan Hamerly, Jeff Hill, Peter McMahon and Amir Safavi-Naeini for helpful discussion.


\bibliographystyle{acm} 
\bibliography{Remote}

\appendix
\section{Basic Component Models}
\label{sec:component_models}

Here we present the component models used to build the perceptron circuit. We will first describe the static components such as beamsplitters, phase shifts and coherent displacements, then proceed to describe the different Kerr-nonlinear models and finally the NOPO model.

\subsection{Static, Linear Circuit Components} 
\label{sub:static_circuit_components}
All of these components have in common that they have no internal dynamics, implying that the $A, B$ and $C$ matrices and the $a$-vector have zero elements, and $A_{\rm NL}$ is not defined.

\subsubsection{Constant Laser Source} 
\label{ssub:constant_laser_source}
The simplest possible static component is given by single input/output coherent displacement with coherent amplitude $\eta$. This model is employed to realize static coherent input amplitudes.
The $D$ matrix is trivially given by $D=(1)$ and the coherent amplitude is encoded in $c=(\eta)$.
This leads to the desired input-output relationship $\beta_{\rm out} = \eta + \beta_{\rm in}$.
For completeness we also provide the SLH \cite{Gough2009Series} model $((1), (\eta), 0)$.

\subsubsection{Static Phase Shifter} 
\label{ssub:static_phase_shifter}
The static single input/outputs phase shifter has $D=(e^{i\phi})$ and $c = (0)$, leading to an input output relationship of $\beta_{\rm out} = e^{i\phi} \beta_{\rm in}$.
Its SLH model is $((e^{i\phi}), (0), 0)$.

\subsubsection{Beamsplitter} 
\label{ssub:beamsplitter}
The static beamsplitter mixes (at least) two input fields and can be parametrized by a mixing angle $\theta$. It has $D = \begin{pmatrix} \cos\theta  & -\sin\theta \\ \sin\theta & \cos \theta \end{pmatrix}$ and $c = (0,0)^T$.
This leads to an input output relationship
\begin{align}
    \begin{pmatrix} \beta_{out,1}\\\beta_{out,2}\end{pmatrix} = \begin{pmatrix} \cos\theta  & -\sin\theta \\ \sin\theta & \cos \theta \end{pmatrix} \begin{pmatrix} \beta_{in,1}\\\beta_{in,2}\end{pmatrix}
\end{align}

Its SLH model is $\left(\begin{pmatrix} \cos\theta  & -\sin\theta \\ \sin\theta & \cos \theta \end{pmatrix}, \begin{pmatrix} 0\\0 \end{pmatrix}, 0 \right)$.

\subsection{Resonator Models} 
\label{sub:resonator_models}
We consider resonator models with $m$ internal modes and $n$ external inputs and outputs. We assume for simplicity that $a = \mathbf{0}$ and $c = \mathbf{0}$ meaning that we will model all coherent displacements explicitly in the fashion described above. We also assume that their scattering matrices are trivially given by $D = \mathbf{1}_n$ which means that far off-resonant input fields are simply reflected without a phase shift. 
Furthermore, none of our assumed models feature \emph{linear} coupling between the internal cavity modes. This implies that the $A$-matrix is always diagonal. We are always working in a rotating frame. 

\subsubsection{Single mode Kerr-nonlinear Resonator} 
\label{ssub:single_mode_kerr}
A Kerr-nonlinearity is modeled by the nonlinear term $A_{\rm NL}^{\rm Kerr}(\alpha) = -i \chi |\alpha|^2 \alpha$ which can be understood as an intensity dependent detuning.
The $A$-matrix is given by $(-\frac{\kappa_T}{2}-i\Delta),$ its $B$-matrix is $-(\sqrt{\kappa_1}, \sqrt{\kappa_2}, \dots, \sqrt{\kappa_n})$, where the total line width is given by $\sum_{j=1}^n\kappa_j = \kappa_T$ and the cavity detuning from any external drive is given by $\Delta$. The $C$-matrix is given by $C=-B^T$.
The corresponding SLH model is 
\begin{align}
  \left(\mathbf{1}_n, \begin{pmatrix} \sqrt{\kappa_1} a \\ \vdots \\ \sqrt{\kappa_n} a\end{pmatrix}, \tilde{\Delta} a^\dagger a + \frac{\chi}{2} a^{2\dagger} a^2\right),
\end{align}
where the detuning differs slightly $\tilde{\Delta} = \Delta + \chi$ as can be shown in the derivation of the Wigner-formalism. \cite{Santori2014Quantum}

The special case of a single mirror with coupling rate $\kappa$ and negligible internal losses is of interest for construsting the phase sensitive amplifier described in Section \ref{sec:variable_gain_amplifiers}. Considering again an input given by a large static bias and a small signal $\epsilon=\frac{1}{\sqrt{2}} (\epsilon_0 +\delta\epsilon)$, the steady state reflected amplitude is to first order 
\begin{align}    
\epsilon'\approx\frac{1}{\sqrt{2}} \left[\eta\epsilon_0 + g_-(\epsilon_0) \delta\epsilon + g_+(\epsilon_0) \delta\epsilon^\ast\right]. 
\end{align}

For negligible internal losses we can give provide exact expressions for $\eta, g_+$ and $g_-$. Rather than parametrizing these by the bias $\epsilon_0$ we parametrize them by the mean coherent intra-cavity amplitude $\alpha_0$. When the system is not bi-stable (see below) relationship \eqref{eq:ssamp_bias} defines a one-to-one map between $\epsilon_0$ and $\alpha_0.$ 
\begin{align}
    \eta & = -\frac{\kappa/2 - i(\Delta + \chi |\alpha_0|^2)}{\kappa/2 + i(\Delta + \chi |\alpha_0|^2)} \quad \Rightarrow |\eta| = 1,\\
    g_- &= 1+\frac{\kappa\left[-\frac{\kappa}{2}+i\Delta+2i\chi |\alpha_0|^{2}\right]}{\left(\frac{\kappa}{2}\right)^2 + \left(\Delta + 2\chi |\alpha_0|^2\right)^2 - |\chi|^2|\alpha_0|^4}, \\
    g_+ &= \frac{i\kappa \chi \alpha_0^2}{\left(\frac{\kappa}{2}\right)^2 + \left(\Delta + 2\chi |\alpha_0|^2\right)^2 - |\chi|^2|\alpha_0|^4},\\
    \epsilon_0 & =  - \frac{1}{\sqrt \kappa}\left[\frac{\kappa}{2} + i(\Delta + i\chi |\alpha_0|^2)\right]\alpha_0. \label{eq:ssamp_bias}
\end{align}

The Kerr cavity exhibits bistability for a particular interval of bias amplitudes if and only if $\Delta/\chi < 0$ and $|\Delta| \ge \frac{\sqrt{3}\kappa}{2}=\Delta_{\rm th}.$

At any fixed bias amplitude and corresponding internal steady state mode amplitude the maximal gain experienced by a small signal is given by $g^{\rm max} = |g_-|+|g_+|$. Here maximal means that we maximize over all possible signal input phases relative to the bias input.
To experience this gain, the signal has to be in an appropriate quadrature defined by $\arg{\delta \epsilon} = \frac{\arg g_- - \arg g_+}{2}.$ The orthogonal quadrature is then maximally de-amplified by a gain of $||g_-|-|g_+||$ and it is possible to show that for negligible losses the perfect squeezing relationship $\left(|g_-|+|g_+|\right)||g_-|-|g_+|| = \left| |g_-|^2 - |g_+|^2\right| = 1$ holds for any bias amplitude. Furthermore, for fixed cavity parameters $g^{\rm max}$ is maximized at a particular  non-zero intra-cavity photon amplitude
\begin{align}\label{eq:n_of_Delta}
  |\alpha_0^{\rm max}|^2 &= \sqrt{\frac{\Delta^2 + \frac{\kappa^2}{4}}{3\chi^2}} \\
  \Rightarrow g^{\rm max}& = \sqrt{\frac{\sqrt{f} + \kappa}{\sqrt{f} - \kappa}},
  \text{ with } f  =  28\Delta^2 + 4\kappa^2 - 8 \Delta \sqrt{12 \Delta^2 + 3 \kappa^2}.
\end{align}
Note that the maximal gain does not depend on the strength of the non-linearity. The relationship between $g^{\rm max}$ and $\Delta$ can be inverted:
\begin{align}\label{eq:delta_of_g}
  \Delta = \frac{\sqrt{3}\kappa}{2} \frac{\left(g^{\rm max}-\sqrt{3}\right)\left(g^{\rm max}-\frac{1}{\sqrt{3}}\right)}{{g^{\rm max}}^2-1}
\end{align}

Using all this it is straightforward to construct a tunable Kerr-amplifier. The symmetric construction proposed in Section \ref{sec:variable_gain_amplifiers} provides the additional advantage that one does not have to cancel the scattered bias.
It is also convenient to prepend and append phase shifters to the signal input and output that ensure $g_-=g_+ = g^{\rm max}/2$ at maximum gain.

The quadrature filter construction relies on the presence of additional cavity losses that are equal to the input coupler $\kappa_2 = \kappa_1 = \kappa.$ In this case the gain coefficients for reflection of the first port are given by
\begin{align}
    g_- &= 1+\frac{\kappa\left[-\kappa+i\Delta+2i\chi |\alpha_0|^{2}\right]}{\kappa^2 + \left(\Delta + 2\chi |\alpha_0|^2\right)^2 - |\chi|^2|\alpha_0|^4}, \\
    g_+ &= \frac{i\kappa \chi \alpha_0^2}{\kappa^2 + \left(\Delta + 2\chi |\alpha_0|^2\right)^2 - |\chi|^2|\alpha_0|^4},\\
    \epsilon_0 & =  - \frac{1}{\sqrt \kappa}\left[{\kappa} + i(\Delta + i\chi |\alpha_0|^2)\right]\alpha_0. \label{eq:ssamp_bias}
\end{align}
and one may easily verify that for dynamic resonance, i.e., $\chi|\alpha_0|^2 = -\Delta$, the gain coefficients are equal in magnitude $|g_-|=|g_+|$ which implies that there exists an input phase for which the reflected signal vanishes.

\subsubsection{Two mode Kerr-nonlinear resonator} 
\label{ssub:two_mode_kerr}
We label the mode amplitudes as $\alpha_1$ and $\alpha_2$. In this case the nonlinearity includes a cross-mode induced detuning 
\begin{align}
    A_{\rm NL}^{\rm Kerr2}(\alpha) = \begin{pmatrix} -i \chi_a |\alpha_1|^2 \alpha_1 - i \chi_{ab} |\alpha_2|^2 \alpha_1  \\  -i \chi_{ab} |\alpha_1|^2 \alpha_2 - i \chi_{b} |\alpha_2|^2 \alpha_2 \end{pmatrix}
\end{align}

The model matrices are
\begin{align}
    A & = \begin{pmatrix}
      -\frac{\kappa_{a,T}}{2}-i\Delta_a &  0 \\
      0 & -\frac{\kappa_{b,T}}{2}-i\Delta_b
      \end{pmatrix}, \\
    B & = -\begin{pmatrix}
        \sqrt{\kappa_{a,1}}& \sqrt{\kappa_{a,2}}&  \dots & \sqrt{\kappa_{a,n_a}} & 0 & \dots & 0 \\
        0 & 0& \dots & 0 & \sqrt{\kappa_{b,1}}& \sqrt{\kappa_{b,2}}&  \dots & \sqrt{\kappa_{b,n_b}}
      \end{pmatrix},\\
    C &= -B^T,
\end{align}
and the corresponding SLH model is
\begin{align}
    \left(\mathbf{1}_{n_a+n_b}, C \begin{pmatrix}
      a \\ b
    \end{pmatrix}, \tilde{\Delta}_a a^\dagger a + \tilde{\Delta}_b b^\dagger b + \frac{\chi_a}{2}a^{2\dagger} a^2 + \frac{\chi_b}{2}b^{2\dagger} b^2 + \chi_{ab}a^\dagger a b^\dagger b\right),
\end{align}
with $\tilde{\Delta}_{a/b} = \Delta_{a/b} + \chi_{a/b} + \frac{\chi_{ab}}{2}$ and where the Wigner-correspondence\footnote{In this appendix we denote expectations with respect to the Wigner function as $\langle \cdot \rangle _{\rm W}$ and quantum mechanical expectations as $\langle \cdot \rangle$.} is $\langle \alpha_1\rangle_{\rm W} = \langle a \rangle$, $\langle \alpha_2\rangle_{\rm W} = \langle b \rangle$.

We briefly summarize how to construct a controlled phase shifter using an ideal two-mode Kerr cavity with a single input coupling to each mode and negligible additional internal losses. We exploit that in this case the reflected steady state signal amplitude $\zeta'$ is identical to the input amplitude $\zeta$ up to a power dependent phase shift
\begin{align}
    \zeta' = -\frac{\frac{\kappa_a}{2} - i \left(\Delta_a + i\chi_{a} |\alpha_0|^2+ i\chi_{ab} |\beta_0|^2  \right)}{\frac{\kappa_a}{2} + i \left(\Delta_a + i\chi_{a} |\alpha_0|^2+ i\chi_{ab} |\beta_0|^2  \right)} \zeta\quad \Rightarrow |\zeta'| = |\zeta|.
\end{align}
We assume that the control input amplitude takes on two discrete values $\xi = 0$ or $\xi = \xi_0$ and that variations of the signal input amplitude are small $|\zeta|\approx |\zeta_0|$. In this case a good choice of detunings and coupling rates is given by
\begin{align}\label{eq:modulator_params1}
    \Delta_a &= \frac{\kappa_a}{2} - \frac{2\chi_a |\zeta_0|^2}{\kappa_a}  \\ 
    \Delta_b &= \frac{\kappa_a\chi_b}{\chi_{ab}} - \frac{2\chi_{ab}|\zeta_0|^2}{\kappa_a} \\
    \xi_0 &= \frac{\sqrt{\kappa_a\kappa_b}}{2\sqrt{|\chi_{ab}|}} \label{eq:modulator_params3}
\end{align}
in addition to two inequality constraints
\begin{align}\label{eq:modulator_constraints}
  \Delta_a &  \le \sqrt{3}\frac{\kappa_a}{2}\\
  \Delta_b & \le \sqrt{3}\frac{\kappa_b}{2}
\end{align}
that ensure that the system is stable. 
This construction ensures that $\frac{\left.\zeta'\right|_{\xi=\xi_0}}{\left.\zeta'\right|_{\xi=0}} = -1$ and in fact it can easily be generalized to the more realistic case of non-negligible internal losses.

Finally note that the inequality constraints imply that the lower bounds for the input couplings scale as $\kappa_a^{\rm min}, \kappa_b^{\rm min} \propto |\zeta_0|$ which is important for our power analysis in Section \ref{sec:power_budget_and_energy_scalings}.
This, in turn implies that $\xi_0 \propto |\zeta_0|$ which is a fairly intuitive result.

The controlled phase shifter can now be included in one arm of a Mach-Zehnder interferometer to create a Fredkin gate (cf.~Section \ref{sec:optical_switches}). 

To realize a thresholder, the control mode input is prepended with a two port Kerr-cavity with parameters chosen such that it becomes dynamically resonant with maximal differential transmission gain close to where its output gives the correct high control input $\xi_0.$ 

Overall, we remark that even when we account for the prepended cavity, the relationship $c \propto |\zeta_0|$ still holds, where $c$ is the input to the thresholder. To see how the total decay rate of the thresholding cavity $\kappa_{\rm thresh}$ scales consider first that to get maximum differential gain or contrast, we ought pick a detuning right at or below the Kerr stability threshold $\Delta \approx \Delta_{\rm th} = \sqrt{3}\kappa_{\rm thresh}/2.$ 

We choose the maximum input amplitude such that it approximately achieves dynamic resonance within the prepended thresholding cavity. This occurs when $\Delta = -\chi |\alpha_0|^2$ (cf.~Appendix \ref{ssub:single_mode_kerr}) and at an input amplitude of $c \propto \sqrt{\kappa_{\rm thresh}\left|\frac{\Delta}{\chi}\right|} \propto \kappa_{\rm thresh}.$

\subsubsection{NOPO model} 
\label{ssub:nopo_model}
The NOPO model has consists of three modes, the signal and idler modes $\alpha_s, \alpha_i$ and the pump mode $\alpha_p$. We assume a triply resonant model\footnote{It is possible to drop this resonance assumption for the pump.} and that $\omega_s + \omega_i = \omega_p$, allowing for resonant conversion of pump photons into pairs of signal and idler photons and vice versa.
The nonlinearity is given by
\begin{align}
    A_{\rm NL}^{\rm NOPO}(\alpha) = \begin{pmatrix}
        \chi \alpha_i^\ast \alpha_p \\
        \chi \alpha_s^\ast \alpha_p \\
        - \chi \alpha_s \alpha_i
    \end{pmatrix}
\end{align}
and the model matrices are
\begin{align}
    A &= \begin{pmatrix}
      -\frac{\kappa}{2} & 0 & 0 \\
      0 & -\frac{\kappa}{2} & 0 \\
      0 & 0 & -\frac{\kappa_{p}}{2}
    \end{pmatrix},\quad
    B = -\begin{pmatrix}
      \sqrt{\kappa} & 0 & 0 \\
      0 & \sqrt{\kappa} & 0 \\
      0 & 0 & \sqrt{\kappa_{p}}
    \end{pmatrix},\\
    C &= -B^T.
\end{align}
Here, the SLH model is given by
\begin{align}
    \left(\mathbf{1}_3, C \begin{pmatrix}
    a \\ b \\ c
    \end{pmatrix}, i\chi\left(abc^\dagger - a^\dagger b^\dagger c\right)\right)
\end{align}
where now $a,b$ and $c$ correspond to $\alpha_s, \alpha_i$ and $\alpha_p$.

A steady state analysis of the system driven only by a pump input amplitude $\epsilon$ reveals that below a critical threshold $|\epsilon| < \epsilon_{\rm th} = \frac{\kappa \sqrt{\kappa_p}}{4 \chi}$ the system as a unique fixpoint with $\alpha_s=\alpha_i=0$ and $\alpha_p = -\frac{2\epsilon}{\sqrt{\kappa_p}}.$
Above threshold $|\epsilon| \ge \epsilon_{\rm th}$, the intra-cavity pump amplitude stays constant at the threshold value $\alpha_p = -\frac{2\epsilon_{\rm th}\epsilon/|\epsilon|}{\sqrt{\kappa_p}} = -\frac{\kappa\epsilon/|\epsilon|}{2\chi}$ and the signal and idler mode obtain non-zero magnitude
\begin{align}
    |\alpha_s| = |\alpha_i| = \sqrt{\frac{4\epsilon_{\rm th}}{\kappa} \left(|\epsilon| - \epsilon_{\rm th}\right)}.
\end{align}
As an interesting consequence of the model's symmetry there exists not a single above threshold state but a whole manifold of fixpoints parametrized by a correlated signal and idler phase
\begin{align}
    \alpha_s & = \sqrt{\frac{4\epsilon_{\rm th}}{\kappa} \left(|\epsilon| - \epsilon_{\rm th}\right)} e^{i\phi + i \phi_0}\\
    \alpha_i & = \sqrt{\frac{4\epsilon_{\rm th}}{\kappa} \left(|\epsilon| - \epsilon_{\rm th}\right)} e^{-i\phi+ i \phi_0}
\end{align}
where the common phase $\phi_0$ is fixed by the pump input phase via
\begin{align}
    \alpha_s\alpha_i = -\frac{4\epsilon_{\rm th}}{\kappa}\left(|\epsilon| - \epsilon_{\rm th}\right) \frac{\epsilon}{|\epsilon|}.
\end{align}
In particular, for $\epsilon < 0$ we have $\alpha_i = \alpha_s^\ast.$
Above threshold the system will rapidly converge to a fixpoint of well-defined phase $\phi$. Without quantum shot noise $\phi$ would remain constant. With noise, however, the system can freely diffuse along the manifold. When the pump bias input is sufficiently large compared to threshold and consequently there are many signal and idler photons present in the cavity at any given time $(|\alpha_{s/i}|^2 \gg 1)$ one can analyze the dynamics along the manifold and of small orthogonal deviations from the manifold. In the symmetric case considered here where signal and idler have equal decay rates, the differential phase degree of freedom $\phi = \frac{\arg \alpha_i - \arg \alpha_s}{2}$ decouples from all other variables and approximately obeys the SDE
\begin{align}
    d\phi &  = \sqrt{\gamma_\phi} dW_t, \quad dW_t^2 = dt\\
\text{with } \gamma_\phi&  = \frac{\kappa}{8|\alpha_s|^2} = \frac{\kappa^2}{32\epsilon_{\rm th} \left(|\epsilon| - \epsilon_{\rm th}\right)}.
\end{align}
It is relatively straightforward to generalize these results to a less symmetric model with different signal and idler couplings and even non-zero detunings, but for a given nonlinearity the model considered here provides the smallest phase diffusion and thus the best analog memory. For a very thorough analysis of this model we refer to \cite{Graham1968QuantumFluctuations}.

\subsection{Composite component models} 
\label{sub:composite_component_models}

Due to the scope of this article, we will refrain from including the full net lists for the composite component models in this article and instead publish them online at \cite{Tezak2014PerceptronFiles}.
We remark that composing a photonic circuit from the above described non-linear photonic models is often complicated by the fact that the steady state input-output relationships are hard or even impossible to invert analytically. A systematic approach to optimizing component parameters would be highly desirable.

\end{document}